\documentclass[%
 reprint,
 amsmath,amssymb,
aps,
prb,
floatfix,
]{revtex4-2}

\usepackage{graphicx}
\usepackage{xcolor}
\usepackage{amsmath}
\usepackage[mathlines]{lineno}

\begin{document}

\title{Orbital-selective Mott phase and spin nematicity\\ in Ni-substituted FeTe$_{0.65}$Se$_{0.35}$ single crystals}

\author{Marta Z. Cieplak$^{1}$, I. Zajcewa$^{1}$, A. Lynnyk$^{1}$, K. M. Kosyl$^{1}$, and D. J. Gawryluk$^{1,2}$}

\affiliation{$^{1}$Institute of Physics, Polish Academy of Sciences, 02 668 Warsaw, Poland\\
$^{2}$Laboratory for Multiscale Material Experiments, Paul Scherrer Institute, 5232 Villigen PSI, Switzerland}

\date{\today}

\begin{abstract}
{The normal state in iron chalcogenides is metallic but highly unusual, with orbital and spin degrees of freedom partially itinerant or localized depending on temperature, leading to many unusual features. In this work, we report on the observations of two of such features, the orbital selective Mott phase (OSMP) and spin nematicity, evidenced in magnetization and magnetotransport [resistivity, Hall effect, anisotropic magnetoresistance (AMR)] of Fe$_{1-y}$Ni$_y$Te$_{0.65}$Se$_{0.35}$ single crystals, with $0 < y < 0.21$. Substitution of Ni dopes crystals with electrons, what eliminates some of the hole pockets from Fermi level, leaving only one, originating from $d_{xy}$ orbital. This leads to electron-dominated conduction at low $T$ for $y \gtrsim 0.06$. However, at high temperatures, $T \gtrsim 125 \div 178$ K, the conduction reverses to hole-dominated. Anomalies in magnetization and resistivity are observed at temperatures which approach high-$T$ boundary of the electron-dominated region. Analysis of these effects suggests a link with the appearance of the $d_{z^2}$ hole pockets at X points of the Brillouin zone in the OSMP phase, facilitated by the localization of $d_{xy}$ orbital, as recently reported by angular resolved photoemission experiments (\textit{J. Huang et al., Commun. Phys. \textbf{5}, 29 (2022)}). The low-$T$ AMR shows mixed 4-fold and 2-fold rotational symmetry of in-plane magnetocrystalline anisotropy, with the 4-fold term the largest at small $y$, and suppressed at intermediate $y$. These results are consistent with the mixed stripe/bicollinear magnetic correlations at small $y$, and suppression of stripe correlations at intermediate $y$, indicating development of spin nematicity with increasing Ni doping, which possibly contributes to the suppression of superconductivity.}
\end{abstract}

\maketitle

\section{Introduction}

While there is a broad agreement that the electron-phonon coupling cannot account for superconducting (SC) transition temperatures ($T_c$'s) in iron-based superconductors (IBS), the nature of the normal state, and it's implications for superconductivity, are still not well understood \cite{Mazin2008,Scalapino2012,Dai2015,Fernandes2017,Fernandes2022}. The normal state is metallic but highly unusual, with orbital and spin degrees of freedom partially itinerant or localized depending on temperature, what appears to be a result of multi-orbital character with important role of the Hund's interaction \cite{Fernandes2022}. This leads to many unusual features observed experimentally. One of them is so-called orbital differentiation, evidenced by angular resolved photoemission (ARPES) experiments, a phenomenon of different degrees of correlations experienced by different Fe-derived $d$ orbitals, three of which ($d_{xy}$, $d_{xz}$, $d_{yz}$) are closest to Fermi level \cite{Fernandes2022,Yi2017}. A related finding is orbital-selective Mott phase (OSMP), in which $d_{xy}$ orbital with large effective mass undergoes localization on the increase of temperature above about 120 K, while other $d$-orbitals remain itinerant \cite{Fernandes2022,Yi2015,Huang2022}. Both localized and itinerant moments contribute to magnetic properties, leading in many cases to magnetic orderings at low $T$, and universal presence of dynamic magnetic correlations in all IBS materials at higher temperatures \cite{Dai2015,Tranquada2020}. A feature common to all IBS is nematicity, the electronic-driven breaking of rotational symmetry of the crystal, which lifts the degeneracy between $d_{xz}$ and $d_{yz}$ orbitals and possibly involves $d_{xy}$ orbital as well \cite{Fernandes2014,Li2020,Rhodes2022}. While in many of the IBS the transition to nematic phase is identical with tetragonal-to-orthorhombic structural transition, either coinciding with magnetic ordering transition, or preceding it, in other compounds or/and at high temperatures the nematicity persists in the form of short-range fluctuations in the absence of long-range orthorhombic order. The relationship between nematicity and the anisotropy of spin correlations (called spin nematicity) is not yet clear, with recent studies suggesting a possible link between these two effects \cite{Jiang2023}. Theories and experiments \cite{Steffensen2019,ZWang2020,Wiecki2021,Song2011,Allan2013,Baek2016} discuss the possible impact of disorder on the emergence of nematicity.

Here we focus on the properties of one class of the IBS, iron chalcogenide system, FeTe$_{1-x}$Se$_x$. The recent finding of the topological surface states at $x \sim 0.5$ \cite{Wang2015,Zhang2018} stimulates detail studies of this system, despite relatively low maximum $T_c$. Prompted by interest in the relationship between disorder and nematicity, we re-visit the problem of the substitution of impurities into Fe-site in the FeTe$_{1-x}$Se$_x$ crystals, which we have studied in previous years \cite{Gawryluk2011,Bezusyy2014,Bezusyy2015,Cieplak2015}. These studies have been partially hampered by crystal inhomogeneity affecting various properties, as discussed by us \cite{Gawryluk2011,Wittlin2012,Cieplak2015,Sivakov2017} and others \cite{Prokes2015,Hartwig2018}.

In the present study we use Fe$_{1+{\delta}-y}$Ni$_y$Te$_{0.65}$Se$_{0.35}$ crystals with broad range of Ni content, $0 < y < 0.21$, grown with small crystallization rate, what improves homogeneity, so that Fe-excess is small ($\delta \lesssim 0.04$). We evaluate the structure, magnetization, transport (resistivity and Hall effect), and anisotropic magnetoresistance (AMR). The most important result is the observation of the anomalies in magnetization, resistivity, and Hall effect, which we link to the coherent-incoherent electronic state crossover into the OSMP, which occurs on increasing temperature \cite{Yi2015}. This effect becomes clearly evident in strongly Ni-doped crystals, in which some of the hole pockets are eliminated from the Fermi surface, leaving only one, of $d_{xy}$ origin, as recently confirmed by photoemission study performed on our samples \cite{Rosmus2019}. This results in electron-dominated conduction at low $T$ in crystals with $y \gtrsim 0.06$, switching into hole-dominated conduction at $T \gtrsim 125 \div 178$ K. This effect is well explained by the contribution to transport from $d_{z^2}$-derived hole pockets, which appear at the Fermi level due to localization of $d_{xy}$ orbital \cite{Huang2022}. While there were previous suggestions that some magnetotransport properties may be explained by the crossover to the OSMP \cite{Otsuka2019}, complicated behavior of the Hall effect in multiorbital system did not allow for definite conclusions. Thus, our result appears to be the first unambiguous evidence of the OSMP from magnetotransport experiment. We also show that mixed 4-fold and 2-fold rotational symmetry of the AMR, observed at low $y$, is replaced by 2-fold symmetry with increasing $y$, consistent with the development of spin nematicity induced by Ni doping.

Before detail discussion of our experiment it is worth to summarize shortly experimental data accumulated over last few years on the evolution, with the change of $x$, of the orbital, spin and nematic properties in the FeTe$_{1-x}$Se$_x$ system. The relevant ARPES data, on orbital differentiation and the OSMP, have been reported for crystals with $0 \leq x \leq 0.44$ \cite{Yi2015,Yi2017,Huang2022}. Neutron scattering experiments map the change of magnetic properties with $x$, from long-range magnetic order in Fe$_{1+y}$Te, to the absence of a long-range order in FeSe at ambient pressure \cite{Li2009,Chen2009,Zaliznyak2011,Tranquada2020}; in addition, short-range magnetic order is observed at small Se content $x \approx 0.33$ \cite{Bao2009,Khasanov2009}. The evolution of magnetic excitations with $x$ or $T$ is well described by a model of disordered cluster state \cite{Zaliznyak2015}, which assumes different local magnetic ordering within clusters consisting of 4 nearest neighbor (NN) Fe-spins, either ferromagnetic (FM) or antiferromagnetic (AFM), and coupling between clusters by a short-range AFM correlations, characterized by two different wave vectors, either Q=($\pi$,0) (bicollinear) or Q=($\pi$,$\pi$) (stripe). While local FM clusters coupled by bicollinear correlations describe well magnetic excitations in Fe$_{1+y}$Te \cite{Zaliznyak2011}, in crystals with intermediate $x$ the AFM clusters have to be assumed, coupled by stripe correlations at low $T$, and by a mixture of stripe and bicollinear correlations at high $T$ \cite{Zaliznyak2015,Xu2016,Xu2017,Xu2018,Tranquada2020}. It is important to point out that AFM clusters break C$_4$ rotational symmetry, suggesting possible link to nematicity. Finally, the substitution of transition metal elements into Fe-site drives the correlations towards bicollinear pattern, independent of temperature \cite{Tranquada2020}.

This picture is further complicated by an occurrence of the nematic phase. The largest body of data exist for FeSe, for which the transition to static nematic phase is found at 90 K \cite{Watson2015,He2018,Rhodes2021}. Recent nuclear magnetic resonance (NMR) experiment shows that this transition, which normally lifts the degeneracy between $d_{xz}$ and $d_{yz}$ orbitals, affects also $d_{xy}$ orbital; moreover, this orbital is also playing a dominant role in spin susceptibility, pointing to spin-orbital-intertwinned nematicity \cite{Li2020}. The nematic phase in compounds with intermediate $x$ is still a subject of studies. Although the tetragonal-to-orthorhombic transition is suppressed for $x \lesssim 0.5$ \cite{Terao2019}, some low-$T$ broadening of the diffraction peaks has been noted at $x=0.5$ \cite{Horigane2009}, together with change of the $T$-dependence of in-plane lattice parameter \cite{Horigane2009,Xu2016}, what may suggest local structural deformations. Lifting of the degeneracy of $d_{xz}$ and $d_{yz}$ orbitals has been shown by ARPES for crystals with $x=0.5$ \cite{Johnson2015}, and  nematicity has been observed for $x=0.4$ by quasiparticle scattering \cite{Singh2015} and elastoresistance \cite{Kuo2016}. In addition, spin nematicity has been recorded by in-plane AMR for $x=0.39$ \cite{Liu2021}. Aside from static nematicity, dynamic fluctuations exist in the whole range of $x$, as shown by recent elastoresistivity measurements for $0 < x < 0.53$, with symmetry changing with $x$, closely resembling evolution of magnetic correlations \cite{Jiang2023}. Finally, recent study of quasiparticle scattering in several crystals in the vicinity of $x=0.45$ reveals strain-induced static electronic nematicity in nanoscale regions, accompanied by SC suppression in the same regions \cite{Zhao2021}.

In the following we will discuss how our experiment may be understood in view od the above findings.

\section{Experimental details}

The crystals of nominal composition Fe$_{1+{\delta}-y}$Ni$_y$Te$_{0.65}$Se$_{0.35}$ were grown by Bridgman method, as described elsewhere \cite{Gawryluk2011}, using crystallization rate of 1.2 mm/h. The crystals show a perfect mirror-like cleavage plane, suggesting their good crystal quality, furthermore confirmed by small full width at half maximum (FWHM) of 004 diffraction peak, of about 1.35-1.67 arc min, as already reported \cite{Gawryluk2011,Wittlin2012}. This is in contrast to more inferior quality of the crystals produced with higher crystallization rates, in which the FWHM increases by a factor of 3.5 or more, as discussed previously \cite{Gawryluk2011,Wittlin2012,Cieplak2015,Sivakov2017}.

The crystals for this study have been carefully selected by quantitative point analysis, performed by energy-dispersive x-ray (EDX) spectroscopy at many points on each crystal. The average Ni content ($y$) has remained close to nominal, with small dispersion, not exceeding $\Delta y = \pm 0.003$, in samples cut from the same crystal. The actual $y$ values, determined by EDX for each sample studied, are used to label samples. The Se content is $0.35 \pm 0.02$. The average Fe+Ni content has been found to exceed slightly 1, as described in the next section.

For the structural evaluation the crystals were milled and characterized by powder X-ray diffraction (CuK$_{{\alpha}1}$ radiation) using X'Pert PRO Alpha-1 MPD (Panalytical) diffractometer. The investigated 2${\theta}$ range was 10$^{\circ}$ to 100$^{\circ}$, with a recording step size of 0.0167$^{\circ}$. Phase analysis and lattice constants determination were performed with a Rietveld refinement, using FullProf software.

For magnetization and transport measurements the crystals were cleaved, with cleavage plane always perpendicular to the $c$-axis. The in-plane shape was slightly elongated in the direction of one of the in-plane main crystallographic axes, as confirmed by X-ray examination. The transport current $I$ has been applied in this direction, after cutting the sample into rectangular shape; this resulted in approximate  orientation of the $I$ (to within 10-20 deg) with one of the main in-plane axes.

The resistivities, in-plane ($\rho$), and Hall ($\rho_{xy}$), were measured by $dc$ and $ac$ four-probe methods, respectively, using Physical Property Measurement System (Quantum Design), in the temperature range 2 to 300 K, and in magnetic fields up to 9 T. The angle-dependent magnetoresistance (AMR) was measured in the magnetic field rotated in-plane around $c$-axis (azimuthal $\phi$ angle) or in the magnetic field tilted away from the $c$-axis towards current direction (planar $\theta$ angle). The accuracy of the angle is about $\pm 5$ deg. The magnetization was measured using SQUID magnetometer (MPMS-7XL, Quantum Design), in a magnetic field applied in two perpendicular in-plane directions (along and perpendicular to $I$), and along $c$-axis, on warming in zero-field cooled (ZFC), and on cooling in field-cooled (FCC) modes.

\section{Results}

\subsection{Structural evaluation}

\begin{figure}
\centering
\includegraphics[width=8.5cm]{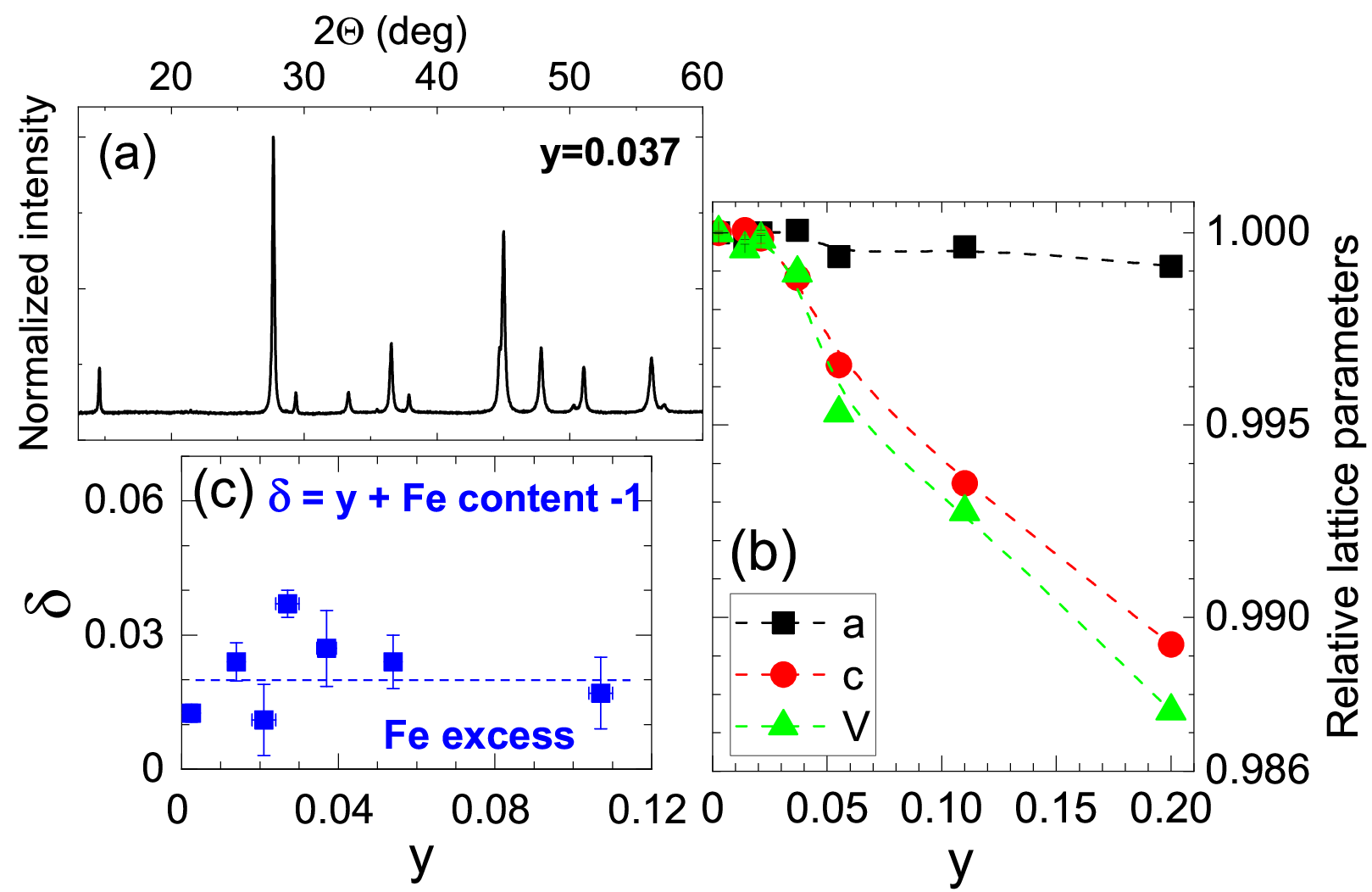}
\caption {(Color online) (a) X-ray diffraction pattern for the crystal with $y$=0.037. (b) Relative lattice parameters vs $y$: $a$ (squares), $c$ (circles), cell volume V (tringles). Errors are smaller than the point size. (c) $\delta$, determined by EDX, versus $y$; the errors are dispersions ${\Delta}{\delta}$ and ${\Delta}y$ determined by EDX.}\label{Sstruc}
\end{figure}

Fig. \ref{Sstruc}(a) shows diffraction pattern for crystal with $y = 0.037$. All diffraction peaks may be indexed by tetragonal phase, with a space group $P4/nmm$. In other crystals tiny traces of Fe$_3$O$_4$ phase, with a space group $Fd$-$3m$, are occasionally seen, with weight fractions below 0.7\%. Fig.\ref{Sstruc}(b) shows the $y$-dependence of the lattice parameters and the cell volume relative to $y = 0$ sample. The lattice parameters change monotonically with Ni doping, in agreement with previous studies \cite{Gawryluk2011}; the $a$ parameter depends very weakly on $y$, while $c$ clearly decreases with increasing $y$; this results in similar decrease of the relative cell volume.

Fig.\ref{Sstruc}(c) shows the deviation from 1 of the total Fe+Ni content, calculated by the relation ${\delta} = y + {\text {Fe content}}-1$, where Fe content and $y$ are measured by EDX. While in principle some of the deviation may occur on the Ni site, very small dispersion of $y$ value suggests that $\delta$ is primarily related to the Fe site. $\delta$ is positive, less then 0.04, pointing to small, almost $y$-independent Fe excess, with the average value of about 0.02. For magnetization and transport measurements the crystals with $\delta < 0.03$ have been selected.

More examples of the diffraction patterns, and discussion of other structural properties, may be found in previous reports \cite{Gawryluk2011,Wittlin2012,Rosmus2019}.

\subsection{Transport properties}

\begin{figure}
\centering
\includegraphics[width=8.5cm]{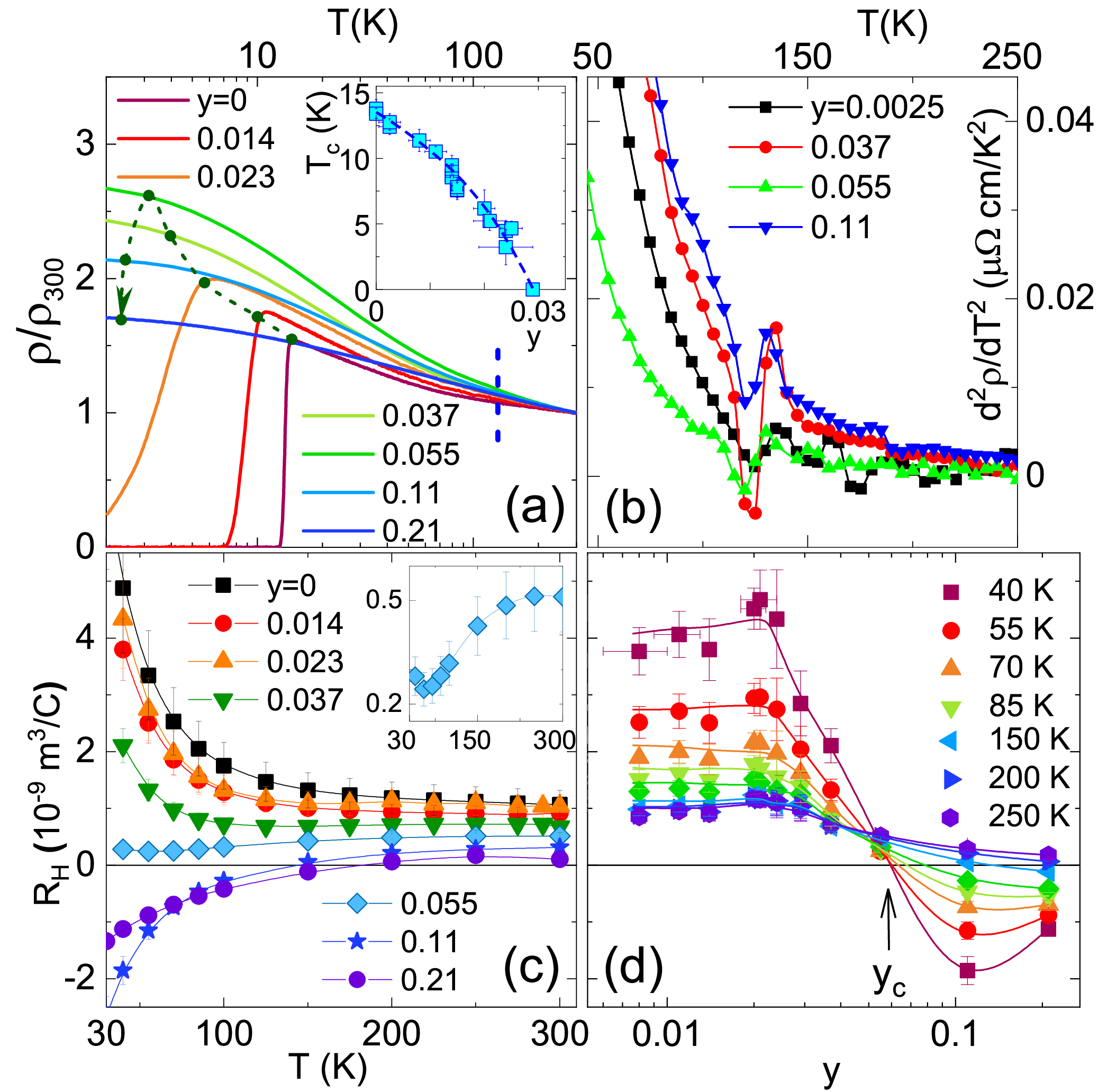}
\caption{(Color online) (a) $\rho /\rho_{300}$ versus $\ln{T}$ for a series of crystals with different $y$. Dashed green arrow illustrates the evolution of the low-$T$ $\rho$ with increasing $y$, and vertical, dashed, blue line marks the temperature, at which the change of slope of $\rho (T)$ occurs. The inset in (a) shows $T_c$ versus $y$. (b) $d^2{\rho}/dT^2$ vs $T$ for several crystals with different $y$. (c) $R_H$ versus $T$ for different $y$. The inset in (c) shows data for y=0.055 sample on expanded scale. (d) $R_H$ versus $\ln{y}$ for different $T$. Arrow indicates $y_c$, at which $R_H$ changes sign at low $T$.}\label{RhoRH}
\end{figure}

The effect of Ni doping on the $T$-dependence of the resistivity is shown in Fig. \ref{RhoRH}(a), where we plot $\rho$, normalized to ${\rho}_{300}$, versus $T$ on a logarithmic scale. Doping introduces $y$-dependent upturn to the resistivity at low $T$, with the magnitude of the upturn well correlated with the suppression of the SC transition temperature, $T_c$. The dependence of the $T_c $ on $y$, depicted in the inset to Fig. \ref{RhoRH}(a), indicates decrease of the $T_c$ to zero at $y \approx 0.03$. Here the $T_c$ is defined as the middle point of the transition, and the vertical error bars reflect 90\% to 10\% transition width.

The dashed green line in Fig. \ref{RhoRH}(a) illustrates non-monotonic evolution of the low-$T$ values of $\rho/\rho_{300}$ with increasing $y$, namely, initial increase, and, after reaching maximum at $y = 0.055$, the decrease. We remark that this does not mean that $\rho$ decreases due to doping, in fact, $\rho_{300}$ at large $y$ is larger by about 20\% from the average $\rho_{300}$ at low $y$, so that resistivity at low $T$ is higher for large doping. Nevertheless, $\rho/\rho_{300}$ rises most steeply with decreasing $T$ at $y=0.055$. The second interesting feature is the change of slope of the resistivity at high temperatures, in the vicinity of 125-130 K, marked in the figure by dashed, vertical, blue line. This change of slope, which is somewhat difficult to see in $\rho (T)$, becomes clearly evident in a form of well-defined anomaly after we calculate the second derivative of the resistivity, $d^2 {\rho}/dT^2$, as depicted in Fig. \ref{RhoRH}(b) for several $y$. The anomaly, seen in the vicinity of 125K, marks clear change of the $T$-dependence of $d^2 {\rho}/dT^2$, which at $T \lesssim 125$ K grows more steeply with decreasing $T$. This feature suggests $T$-induced modification of multi-carrier conduction, the origin of which will be discussed later.

In Fig.\ref{RhoRH}(c) we display $T$-dependence of the Hall coefficient, $R_H$, at different $y$, and in Fig.\ref{RhoRH}(d) the $R_H$ is shown as a function of $\ln{y}$ for different temperatures. The data are extracted from the magnetic field dependence of the Hall resistivity, $\rho_{xy}$, which is linear for $T > 40$ K. We observe that at high temperatures the $R_H$ is positive for all $y$, indicating that hole carries dominate the transport. At lower temperatures the $R_H$ changes sign at $y \gtrsim y_c$, where $y_c \thickapprox 0.06$; this is approximately the same $y$ value at which $\rho/\rho_{300}$ at low $T$ reaches maximum value and starts to decrease. In fact, traces of negative contribution to the $R_H$ are already well visible for $y=0.055$, in the form of downturn of $R_H$, which exceeds experimental errors for temperatures lower then about 150 K [inset to Fig.\ref{RhoRH}(c)]. These effects strongly suggest the electron doping of crystals by Ni substitution.

In order to get a clear confirmation of the electron doping from transport, we have analyzed the Hall resistivity, $\rho_{xy}$, measured at $T \leqq 30 K$. At these temperatures the  $\rho_{xy}$ is nonlinear versus magnetic field, indicating multicarrier conduction. We have extracted the contribution of different types of carriers to transport using matrix formalism, proposed in the past for description of multicarrier semiconductor systems \cite{Kim1993,Kim1999} (see Appendix A for examples of $\rho_{xy} (H)$, and detail description of the fitting procedure). In this formalism, all carriers with the same mobility $\mu_i$ contribute to one type of carriers with concentration $n_i$ ($i=1,2,...$). Here we limit our considerations to 2 types of carriers in each sample ($i=1,2$), because this describes satisfactorily the data; however, one should remember that this is an effective description, i.e. some types of carriers may include possible combination of carriers from different bands with similar mobility.

\begin{figure}
\centering
\includegraphics[width=8.5cm]{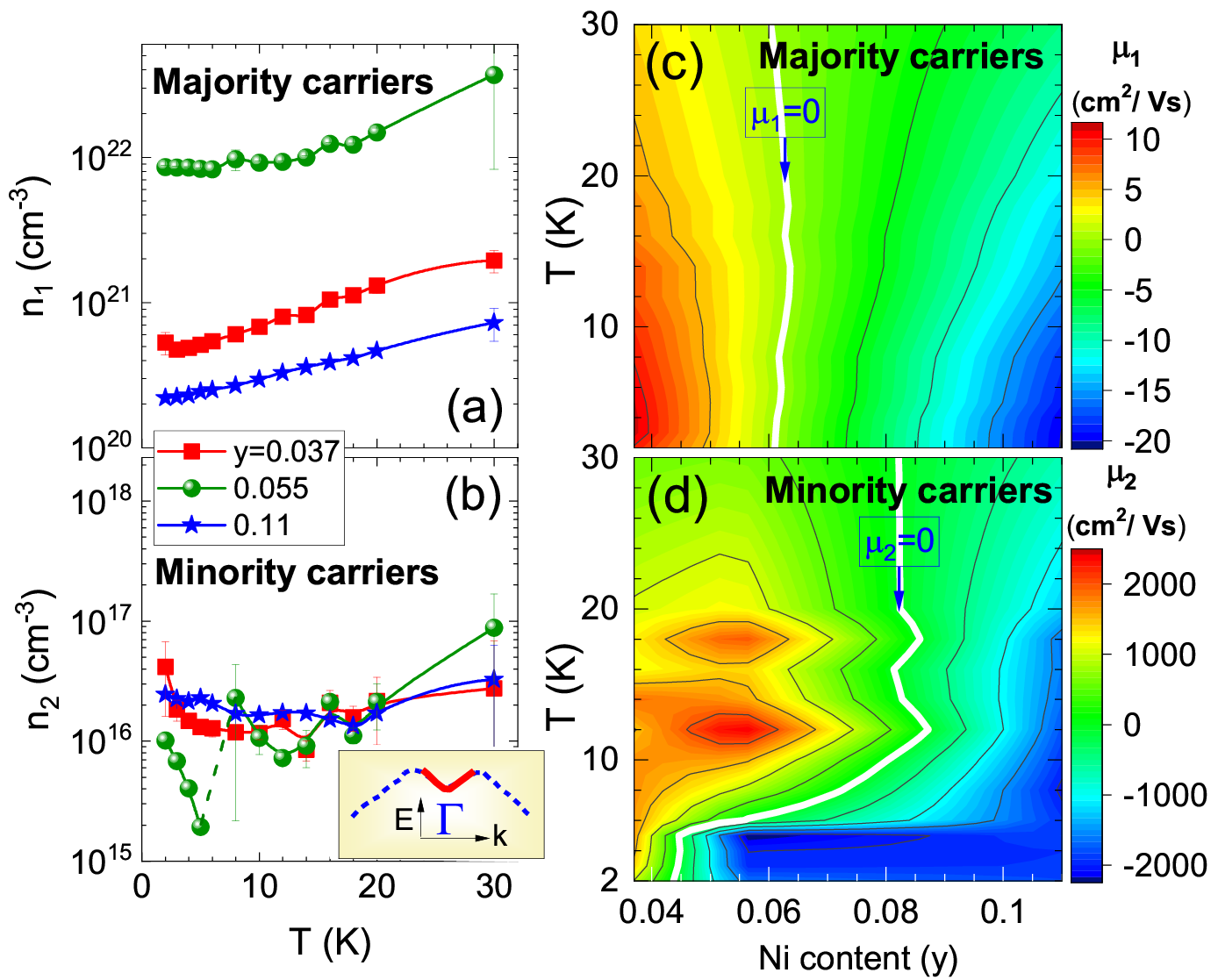}
\caption{(Color online) Carrier concentrations $n$ versus $T$ (a) and (b), and color maps of mobilities $\mu (T,y)$ (c) and (d). In most cases the errors for $n$ are smaller than the point size, whenever it is otherwise, the errors are shown in the figures. The white lines in (c) and (d) indicate zero mobility, when the sign of carriers changes. The inset in (b) shows schematically the possible dispersion around $\Gamma$ point, electron-like (red) and hole-like (blue). }\label{LowHall}
\end{figure}

In each crystal we find majority carriers with larger concentration and smaller mobility ($n_1$ and $\mu_1$), and minority carriers with smaller concentration and larger mobility ($n_2$ and $\mu_2$). Both majority and minority carriers may be positive or negative, depending on the Ni doping. Here, we define positive (negative) mobilities for positive (negative) carriers, and create maps of $\mu_i (T,y)$ shown in Figs. \ref{LowHall} (c) and (d). In the maps positive mobilities (holes) are indicated by red/yellow/light-green colors, while negative mobilities (electrons) are shown by green/blue; the white line shows the boundary between them at $\mu = 0$. In Figs. \ref{LowHall} (a) and (b) we show the $T$-dependence of $n_i$ (on a logarithmic scale) for several selected crystals.

The map of $\mu_1 (T,y)$ for majority carriers [Fig.\ref{LowHall}(c)] shows that they undergo the change of sign from positive at low $y$ to negative at large $y$. This change of sign occurs at $y_c \approx 0.06$, almost independently of temperature. The drop of carrier mobility to zero at $y_c$ explains the strongest increase of $\rho/\rho_{300}$ observed for $y=0.055$ at low $T$.

On the other hand, minority carriers [Fig.\ref{LowHall}(d)] show unusual behavior: they are positive below about $y = 0.045$, and for a substantial $y$ range ($0.045 < y < 0.08$) they remain electron-like at low $T$, changing into hole-like at higher $T$. The abrupt change of carrier polarity is also reflected in the dependence $n_2 (T)$ for crystal with $y=0.055$ [Fig.\ref{LowHall}(b)]. This dependence shows two branches. At $T > 6$ K there is an $n_2$ branch related to positive carriers, first decreasing, and then slightly increasing with decreasing $T$. At $T < 6$ K a second $n_2$ branch appears, increasing with decreasing $T$, related to electron carriers. There are two possible explanations for this unusual behavior. One possibility is that with increasing $y$ and decreasing $T$ the concentration of electrons from M pockets becomes so large, that they are no longer compensated by holes. However, it is unclear why this would lead to change of sign at $y$ lower than $y_c$. Another possibility is that this feature is caused by electrons not from M pockets, but from different band. We note that there are ARPES observations of the electron band at $\Gamma$ point just above Fermi level in FeTe$_{0.6}$Se$_{0.4}$ \cite{Okazaki2014} or in FeTe$_{0.55}$Se$_{0.45}$ surface-deposited by potassium \cite{Zhang2014}, likely originating from antibonding $d_{xy}$ orbital and Se $p_z$ orbital \cite{Zhang2014,Wang2015}. Surface-deposited crystals are doped with electrons, resembling the situation similar to present case of Ni substitution. Electron doping moves upwards the chemical potential, and the bottom of electron band at $\Gamma$ point begins to cross Fermi level, producing peculiar dispersion, which we sketch schematically in the inset in Fig. \ref{LowHall}(b), with electron-like dispersion at small wavevector (shown in red), and hole-like dispersion for larger wavevector (in blue). Such dispersion could result in the electron-like carriers at low $T$, switching into hole-like carriers at larger $T$.

\begin{figure}
\centering
\includegraphics[width=7cm]{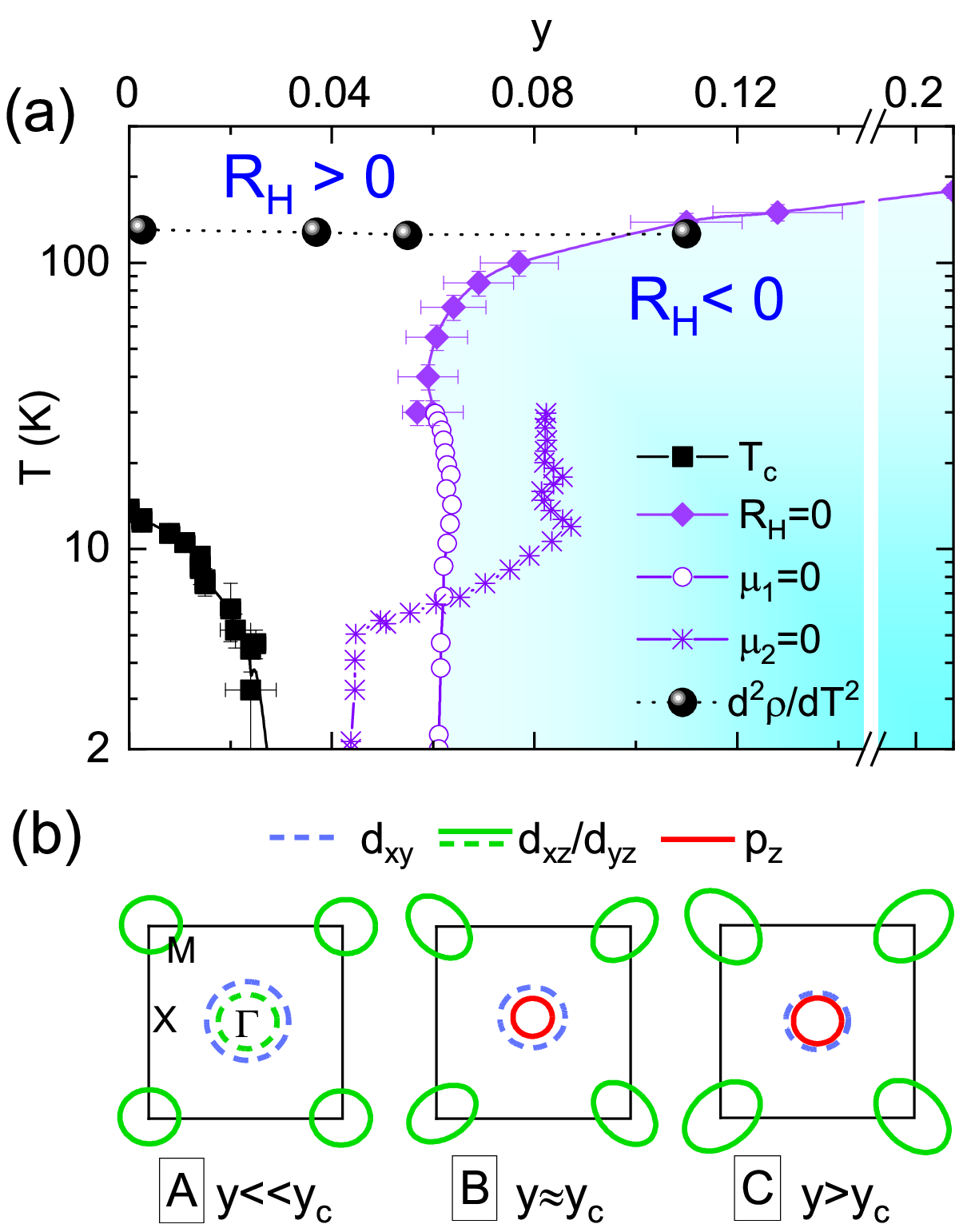}
\caption{(Color online) (a) Phase diagram based on transport experiments. Points indicate temperatures: $T_c$ (squares), $R_H =0$ (diamonds), ${\mu}_1 = 0$ (open circles), ${\mu}_2 = 0$ (stars), resistive anomalies (black spheres). All lines are guides to the eye. Blue background indicates region with dominant electron-like conduction. (b) Schematics of the band structure at the Fermi level at low $T$ for (A) $y \ll y_c$, (B) $y \approx y_c$, and (C) $y > y_c$; continuous and dashed lines indicate electron-like and hole-like bands, respectively.}\label{TDiag}
\end{figure}

We summarize the transport experiments with the phase diagram, shown in Fig. \ref{TDiag}(a), where we plot $T_c (y)$, and the temperatures, at which $R_H = 0$ ($R_H$-line) and mobilities ${\mu}_i = 0$ (${\mu}_i$-lines), as well as the temperature, at which resistive anomaly is seen in $d^2\rho/dT^2$. The most important feature is related to the dependencies on $y$ of the ${\mu}_1$-line, and it's extension to higher temperatures, $R_H$-line. While at low temperature ${\mu}_1$-line runs at almost constant $y_c$ value, the $R_H$-line shifts to higher $y$ with increasing $T$, approaching the region of black spheres, which represent the temperatures of the resistive anomaly. The $R_H$-line eventually saturates, reaching the value of about 178 K at $y=0.21$. Thus, electron carriers dominate transport exclusively in one portion of the phase diagram, marked in the figure by blue color, even in the case of crystals, which are heavily Ni-doped. This is an unusual finding, suggesting that some type of Fermi surface reconstruction occurs at high temperatures, in the vicinity of resistive anomaly, which turns electron-dominated conduction into hole-dominated conduction. A most plausible explanation is the appearance at high-$T$ of the $d_{z^2}$-derived hole pockets at X points of Fermi surface, facilitated by the localization of $d_{xy}$ band in the OSMP phase, as recently uncovered by ARPES experiments in FeTe$_{0.56}$Se$_{0.44}$ \cite{Huang2022}. Note that such interpretation would be consistent with steeper increase of $d^2 {\rho}/dT^2$ with decreasing $T$ below the temperature of the resistive anomaly, because at lower $T$ the $d_{z^2}$-derived holes cannot participate in the transport. In the next sections we will present experiments which support this interpretation.

\begin{figure*}
\centering
\includegraphics[width=16cm]{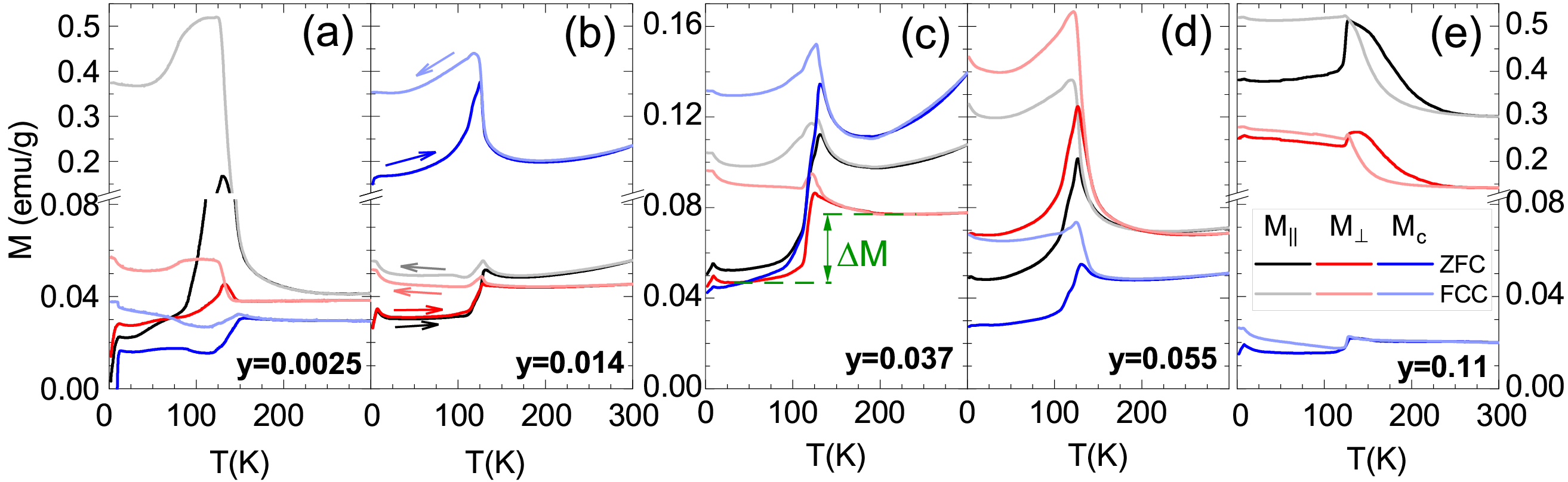}
\caption{(Color online) $T$-dependence of $M$ for crystals with different $y$, 0.0025 (a), 0.014 (b), 0.037 (c), 0.055 (d), 0.11 (e), measured in magnetic field of 100 Oe with different  orientations, $M_{\parallel}$ (black), $M_{\perp}$ (red), and $M_c$ (blue), in ZFC mode (dark lines) and FCC mode (light lines). Note the break of vertical scale in (a), (b) and (e). }\label{Mag}
\end{figure*}

Before turning to other experiments, we mention that the electron doping by Ni substitution has been directly confirmed by recent ARPES measurements, performed at $T = 18$ K for crystals from the same batch as studied here, with $y = 0$, 0.05 and 0.11 \cite{Rosmus2019}. It is found that doping leads to the expansion of the electron pockets at the M point of Brillouin zone, which become elliptical upon doping, and to survival of only one hole pocket at $\Gamma$ point, designated as $\gamma$ pocket, which is derived primarily from $d_{xy}$ orbital \cite{pockets}. Based on this and other ARPES experiments \cite{Rosmus2019,Yi2015,Huang2022}, and on present transport results we illustrate schematically in Fig. \ref{TDiag}(b) the low-$T$ evolution of the band structure at the Fermi level upon doping, from very low $y$ (A), to $y \approx y_c$ (B) and to large electron doping $y > y_c$ (C). In A two hole pockets (of $d_{xy}$ and $d_{xz}/d_{yz}$ origins) cross Fermi level at $\Gamma$ point, and  electron pockets are present at M points. In B electron pockets at M points expand and become elliptical, while at $\Gamma$ point hole pocket $d_{xz}/d_{yz}$ is eliminated and $d_{xy}$ pocket survives, accompanied (hypothetically) by small electron pocket of $p_z$ origin. Finally, in C all electron pockets expand more, while the $d_{xy}$ hole pocket still survives but shrinks.

\subsection{Magnetization}

Now we turn attention to magnetization ($M$), shown in Figs. \ref{Mag}(a-e) for several crystals with different $y$, measured in ZFC and FCC modes in low magnetic field (100 Oe), applied in three mutually perpendicular directions, in-plane, parallel to transport current $I$ ($M_{\parallel}$), in plane, perpendicular to $I$ ($M_{\perp}$), and out-of-plane, parallel to $c$-axis ($M_c$). We observe that all components of $M$ are finite, indicating that the magnetic moments are not aligned strictly in-plane or out-of-plane, but point at some intermediate direction. Significant anisotropy is present, both $y$- and $T$-dependent.

Looking first at low-$T$ data, we observe SC transition in least-doped sample, evident for all field directions [Fig. \ref{Mag}(a)]. On increasing temperature the most prominent feature in the normal-state magnetization appears, that is, the anomaly, present for all $M$ components in the region 125-130 K, in all samples. The anomaly produces substantial difference between low-$T$, and high-$T$ magnetization, accompanied by a peak. The data measured in the FCC mode signalize irreversibility, which is confined to temperatures below the peak in all crystals with $y < 0.06$, but extends to much higher $T$ in highly doped crystal.

Similar $M$ anomaly at temperature of about 125 K has been reported before in some Fe$_{1+y}$Te$_{1-x}$Se$_{x}$ single crystals  \cite{Rossler2010,Fang2008,Wittlin2012} but it has not been studied in detail. It has been speculated that the anomaly may originate from the excess of the Fe \cite{Rossler2010}, or from impurity phases, indicating either Verwey transition in Fe$_3$O$_4$ \cite{Verwey1939,Friedrich2002}, or spin reorientation transition in Fe$_7$(Se-Te)$_8$ impurity \cite{Wittlin2012}. We see no correlation whatsoever between the magnitude of the anomaly and the tiny content of Fe$_3$O$_4$ estimated by X-ray diffraction, and we do not detect Fe$_7$(Se-Te)$_8$ in our samples. Thus, we believe that the anomaly is caused by intrinsic properties of the FeTe$_{1-x}$Se$_x$ system. The fact that the anomaly is observed in the same $T$-region as the resistive anomaly strongly suggests that the two features may be linked.

\begin{figure}
\centering
\includegraphics[width=6.5cm]{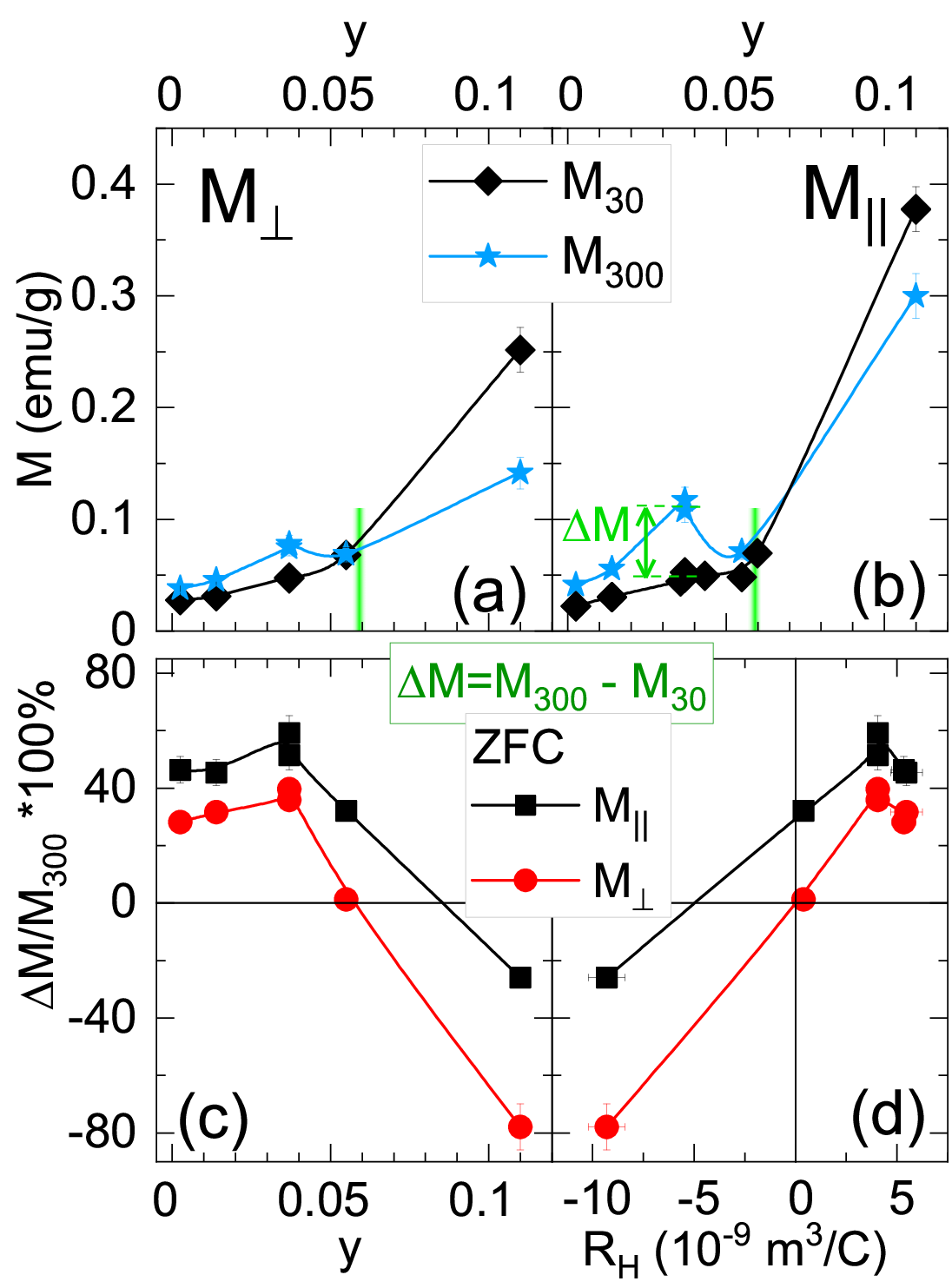}
\caption{(Color online) (a-b) $M_{\perp}$ (a) and $M_{\parallel}$ (b) versus $y$, measured in ZFC mode at $T=30$ K ($M_{30}$, black diamonds) and at $T=300$ K ($M_{300}$, blue stars); green vertical bars show boundary at $R_H = 0$. (c-d) In-plane $\Delta M/M_{300}$, measured for $M_{\parallel}$ (black squares) and for $M_{\perp}$ (red circles) versus $y$ (c), and versus $R_H$ at 30 K (d); definition of $\Delta M$ is shown in Figs. \ref{Mag}(c) and \ref{DeltaM}(b). Unless shown, the errors are smaller than the point size.}\label{DeltaM}
\end{figure}

For further discussion, we extract $M$ measured in ZFC mode at $T = 30$ K (denoted as $M_{30}$) and at $T=300$ K ($M_{300}$). The $y$-dependencies of in-plane components of $M_{30}$ and $M_{300}$, $M_{\perp}$ and $M_{\parallel}$, are shown in Figs. \ref{DeltaM}(a) and \ref{DeltaM}(b), respectively. Both $M_{30}$ components increase monotonically with increasing $y$, reflecting enhanced localization induced by scattering due to Ni doping. The increase is mild, and anisotropy of the in-plane $M_{30}$ is small for $y \lesssim 0.06$, to the left of green vertical bars, which in both figures show $R_H = 0$ boundary, at which low-$T$ conduction changes from hole-dominated into electron-dominated. At $y > 0.06$ the dramatic increase of both components of $M_{30}$ occurs, accompanied by significant anisotropy. Turning to $M_{300}$ components, we observe that they are non-monotonic functions of $y$: at $y \lesssim 0.06$ they are larger than the $M_{30}$ values, but they start to decrease on the approach to $y=0.06$, and they become smaller than the $M_{30}$ values at $y > 0.06$. These dependencies point to a strong link between transport and magnetic properties.

We quantify the difference between low-$T$ and high-$T$ magnetization by defining $\Delta M = M_{300} - M_{30}$, as illustrated in Figs. \ref{Mag}(c) and \ref{DeltaM}(b). $\Delta M$ depends on $y$, and, while in most cases $\Delta M$ is positive, it changes sign into negative for crystal with the largest $y$, as already evident in Fig.\ref{Mag}(e); this is clearly related to enhanced magnetization at low $T$ in this crystal. The $y$-dependence of $\Delta M$, relative to $M_{300}$, is shown in Fig.\ref{DeltaM}(c) for both in-plane directions. The data show slight increase at small $y$, reaching a substantial value of about 40 \%. This is followed by rapid decrease and change of sign at large $y$. There is also anisotropy of $\Delta M/M_{300}$ for two in-plane directions systematically increasing with increasing $y$. Leaving aside for the moment the anisotropy of $\Delta M$, we note that the dependence of $\Delta M$ on $y$ resembles the dependence of Hall coefficient on $y$, shown in Fig.\ref{RhoRH}(d). Therefore, in Fig.\ref{DeltaM}(d) we re-plot the data versus $R_H$ measured at 30 K, on exactly the same piece of crystal as the measurement of $M$. It is evident that quite good proportionality exists between the $R_H$ and $\Delta M/M_{300}$.

The correlation between $\Delta M$ and $R_H$ supports the idea that the positive $\Delta M$ observed in samples with small $y$, i.e. with hole-dominated transport, is a result of delocalization of carriers originating in $d_{xy}$ hole pocket, which occurs on decreasing temperature, in agreement with the observation of the incoherent-coherent transition by ARPES in FeTe$_{0.56}$Se$_{0.44}$ \cite{Yi2015}. This transition reduces localized moment with corresponding increase of the density of itinerant carriers, what explains the reduction of magnetization by almost 40\%. Such huge change, observed in all crystals with hole-dominated transport, cannot be due to tiny (and random) Fe excess, which we find in these crystals. This effect resembles the large change of susceptibility on cooling (by about 30\%) due to delocalization of one of the three Fe electrons in Fe$_{1+\delta}$Te \cite{Li2009,Chen2009,Fobes2014,Tranquada2020}, with one important difference: in case of Fe$_{1+\delta}$Te the change of susceptibility is accompanied by large drop of resistivity at temperatures below the magnetostructural transition. This is not observed here; the only effect in resistivity is small anomaly, visible in second derivative in the vicinity of the peak. This difference suggests strong scattering of carriers in the present case, and, as we will argue, it stems from inhomogeneous magnetic ordering in our mixed crystals.

On the approach to $y \thickapprox 0.6$ we observe substantial reduction of $\Delta M/M_{300}$, well correlated with the $R_H$, followed by a change of sign in electron-doped crystal with large $y$, indicating strong localization on decreasing $T$. Localization in heavily doped crystals is expected, because doping leads to scattering of carriers, what results in the decrease of the density of coherent states and emergence of incoherent states, as predicted by theories \cite{Wadati2010,Berlijn2012}. An example of such effect is strong localization induced by Cu doping in Fe$_{0.98}$Te$_{0.5}$Se$_{0.5}$ crystals leading to metal-insulator transition \cite{Wen2013,Wang2022}. While Cu ion acts as strong scatterer without inducing electron doping, Ni impurity acts both as an electron donor, so that crystals remain metallic, and as a scattering center, inducing localization at sufficiently large doping level. The growing localization with increasing $y$ leads also to enhanced anisotropy of $\Delta M$. We note that the observation of negative $\Delta M$ does not necessarily mean that $d_{xy}$ orbital is not delocalized at low $T$, in fact, it has been very faintly visible in ARPES experiment on our crystal with $y=0.11$ \cite{Rosmus2019}. Rather, it is likely that $d_{xy}$ orbital becomes significantly broadened, so that localizing effect of carrier scattering overcomes the influence of low $T$ delocalization, and magnetization increases on decreasing $T$. On the other hand, at high temperatures scattering-induced localization is diminished, and, in addition, contribution of $d_{z^2}$-derived holes appears at the Fermi level. Both effects lead to the reduction of magnetization.

\begin{figure}
\centering
\includegraphics[width=8cm]{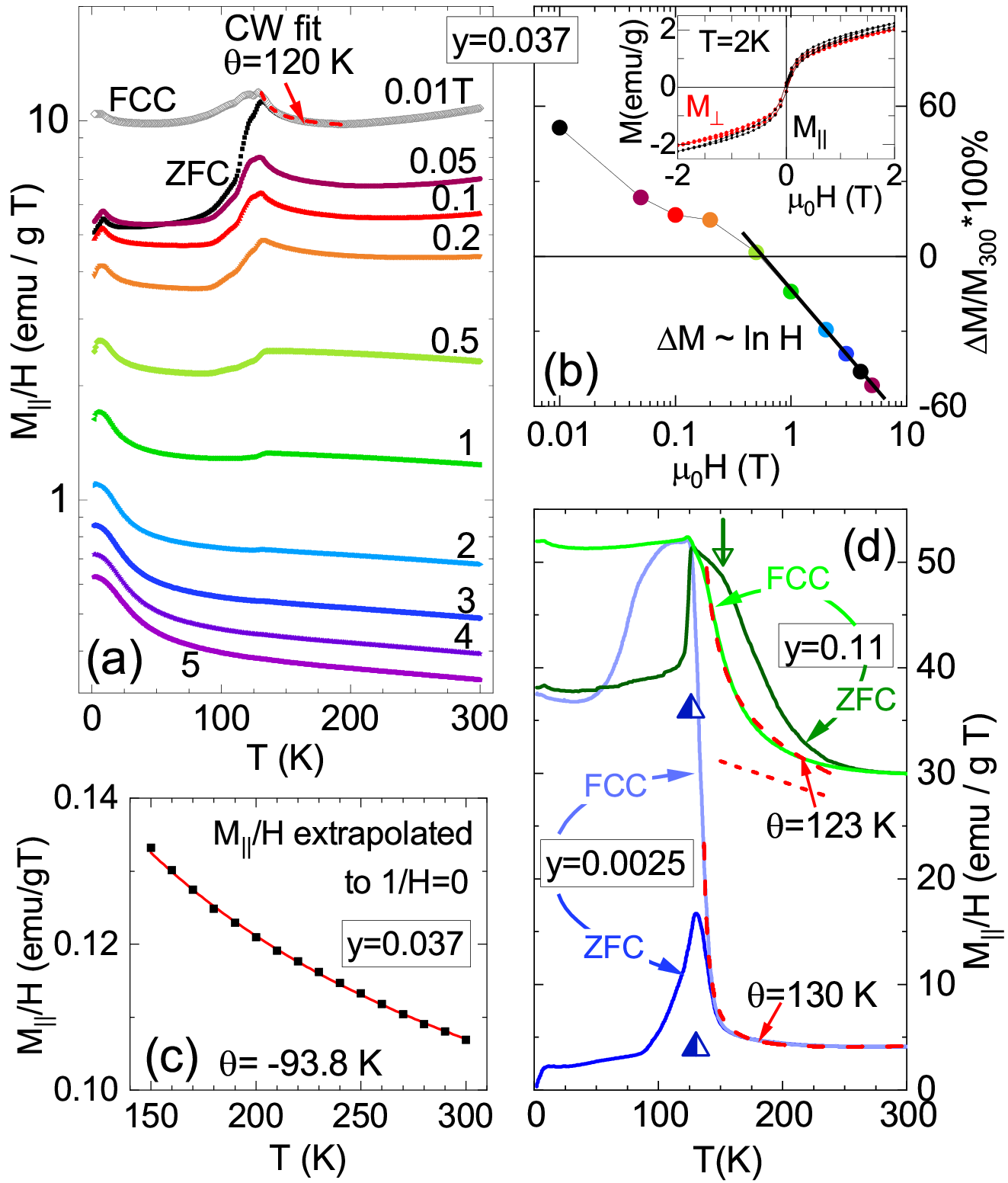}
\caption{(Color online) (a) $M_{\parallel} /H$ (logarithmic scale) versus $T$ for $y = 0.037$, measured in magnetic fields labeled in the figure. (b) $\Delta M/M_{300}$ versus $\ln{\mu_0 H}$; straight line emphasizes proportionality at large field. Inset shows $M-H$ loops for $M_{\parallel}$ and $M_{\perp}$ at $T=2$ K. (c) $M_{\parallel} /H$ vs $T$ extrapolated to $1/H=0$; red line shows CW fit. (d) $M_{\parallel} /H$ for $y=0.0025$ (blue lines) and $y=0.11$ (green lines), measured in ZFC (dark lines) and FCC (light lines); green arrow marks shoulder on ZFC $y=0.11$ curve. In (a) and (d) dashed red lines show CW fits to FCC data; short-dash line shows $\chi_0$ assumed for $y=0.11$. Triangles in (d) indicate positions of resistive anomalies. }\label{Susc}
\end{figure}

In addition to the influence of Ni doping on $M$ anomaly, we also find that anomaly is affected by increasing magnetic field. An example, measured for crystal with $y=0.037$, is presented in Fig. \ref{Susc}(a), where we show $T$-dependence of susceptibility $M_{\parallel}/H$ for magnetic fields in the range from 0.01 to 5 T (ZFC for all fields, and FCC for $\mu_0 H = 0.01$ T; note logarithmic vertical scale). It is seen that the increase of $H$ rapidly suppresses the peak and affects $\Delta M$; while at low magnetic field $\Delta M$ is positive, it changes sign into negative in higher field, reflecting the growth of $M$ at low $T$. $\Delta M/M_{300}$ versus $\mu_0 H$ on logarithmic scale is shown in Fig. \ref{Susc}(b); at $\mu_0 H > 1$ T it follows perfect straight line, $\Delta M/M_{300} \sim \ln{H}$. Such behavior stems from localization, which enhances $M$ at low $T$ in the presence of high magnetic field. The origin is strong scattering of carriers, which is particularly large for this orientation of the field due to anisotropy of magnetic correlations (this is confirmed by magnetoresistance measurements, which we discuss in next section).

Remarkably, the effect of the magnetic field on $\Delta M/M_{300}$ is similar to the effect of Ni doping, both large magnetic field and large doping reverse the sign of $\Delta M$ from positive into negative. However, while magnetic field suppresses the peak, the doping does not. Thus, while the peak and the $\Delta M$ appear together, they seem to be two separate features.

Since magnetic correlations are known to persist in this system, it is natural to consider magnetic phase transition as the origin of the peak feature. While long-range magnetic order is expected to be absent in our crystals, low-$T$ short-range ordering has been documented at low $x$ \cite{Bao2009,Khasanov2009}. Enhanced fluctuations on the approach to transition give rise to anomalies both in the magnetization \cite{Fisher1967}, and in the derivative of resistivity \cite{Fisher1968,nematic}.

The observation of irreversibility indicates that the putative phase transition may have significant FM component. To determine the dominant magnetic interactions at low and at high magnetic fields, we fit susceptibility $\chi = M/H$ measured in FCC mode at high $T$ to Curie-Weiss (CW) relation $\chi = \chi_0 + C/(T-\theta)$. Here $C$ is Curie constant, $\theta$ is the Weiss temperature, and $\chi_0$ includes various $T$-independent contributions to susceptibility. In case of high-field regime we use Honda-Owen method \cite{Honda1910} of extrapolation of the data to the limit $1/H = 0$. In this method $M/H$, measured for high magnetic field (here above 3 T), is plotted versus $1/H$ for any fixed $T$ in the range between 160 K and 300 K, and the value of $M/H$ for $1/H = 0$ is extracted. These values are shown in Fig. \ref{Susc}(c), together with the CW fit, from which we get negative Weiss temperature, $\theta = -93.8$ K, indicating that at high magnetic field the dominant interactions are of AFM-type. Performing similar CW fit at low field is more difficult, because term $\chi_0$ is not constant in the present system; nevertheless, it may be done in narrow region above the peak, as shown for crystal with $y=0.037$ in Fig.\ref{Susc}(a), and for crystals with $y=0.0025$, and 0.11 in Fig.\ref{Susc}(d) (in case of $y=0.11$ we assume $T$-dependent $\chi_0$ shown by short-dashed line). These fits give positive values of $\theta$, indicating FM-type dominant magnetic interactions at low field. This is confirmed by the measurement of $M$-$H$ loops at $T=2$ K (inset to Fig. \ref{Susc}(b)), which shows an approach to saturation above 1 T; extrapolation to $H=0$ gives very small magnetic moment, about 1.3 emu/g (0.04 $\mu_B$/f.u.). Note that both positive and negative values of $\theta$ are small, what indicates strong frustration due to presence of both the AFM and the FM interactions. Thus, while the FM component contributes to the peak, it is unlikely to be a sole contribution.

This conclusion is supported by the observation that the peaks are broad, suggesting inhomogeneous magnetic ordering. This is illustrated in Fig.\ref{Susc}(d), in which we re-plot on the same scale ZFC and FCC components of $M_{\parallel} /H$ for two crystals, $y=0.0025$ and $y=0.11$ (small $M_c$ in these crystals suggests that $M$ is almost in-plane aligned). Temperatures of the resistive anomalies, marked by triangles at the bottom of the peaks, indicate good correlation with the maxima of the peaks. The plot reveals a dramatic difference in the shape of two ZFC peaks. While in least-doped crystal there is a broadening of the peak on the low-$T$ side of the maximum, followed by relatively abrupt decrease of $M_{\parallel}$ on the high-$T$ side, in crystal with $y=0.11$ the inverse behavior is observed, with abrupt increase of $M_{\parallel}$ on low-$T$ side, followed by a shoulder at about 152 K (marked by green arrow). The difference appears also in FCC mode. In least doped crystal the FCC curve deviates upwards from ZFC curve on cooling towards the peak maximum. On the other hand, in case of $y=0.11$ the irreversibility appears around 270 K, and on cooling the FCC curve deviates downwards from the ZFC curve, showing large ($\thicksim 30$ K) thermal hysteresis; this suggests that FM component is confined to the low-$T$ side of the peak, in close vicinity of the resistive anomaly. Interestingly, the FCC curves rise on cooling to almost the same magnitudes for both crystals suggesting that the FM components are similar in both cases; the difference appears on further cooling towards $T=0$, when the $M_{\parallel} /H$ stays saturated at high value in case of $y=0.11$, but decreases for $y=0.0025$. This difference on cooling reflects the degree of localization: while in $y=0.11$ crystal localized moment remains high at low $T$, in $y=0.0025$ crystal localized moment is reduced on cooling.

\begin{figure}
\centering
\includegraphics[width=8cm]{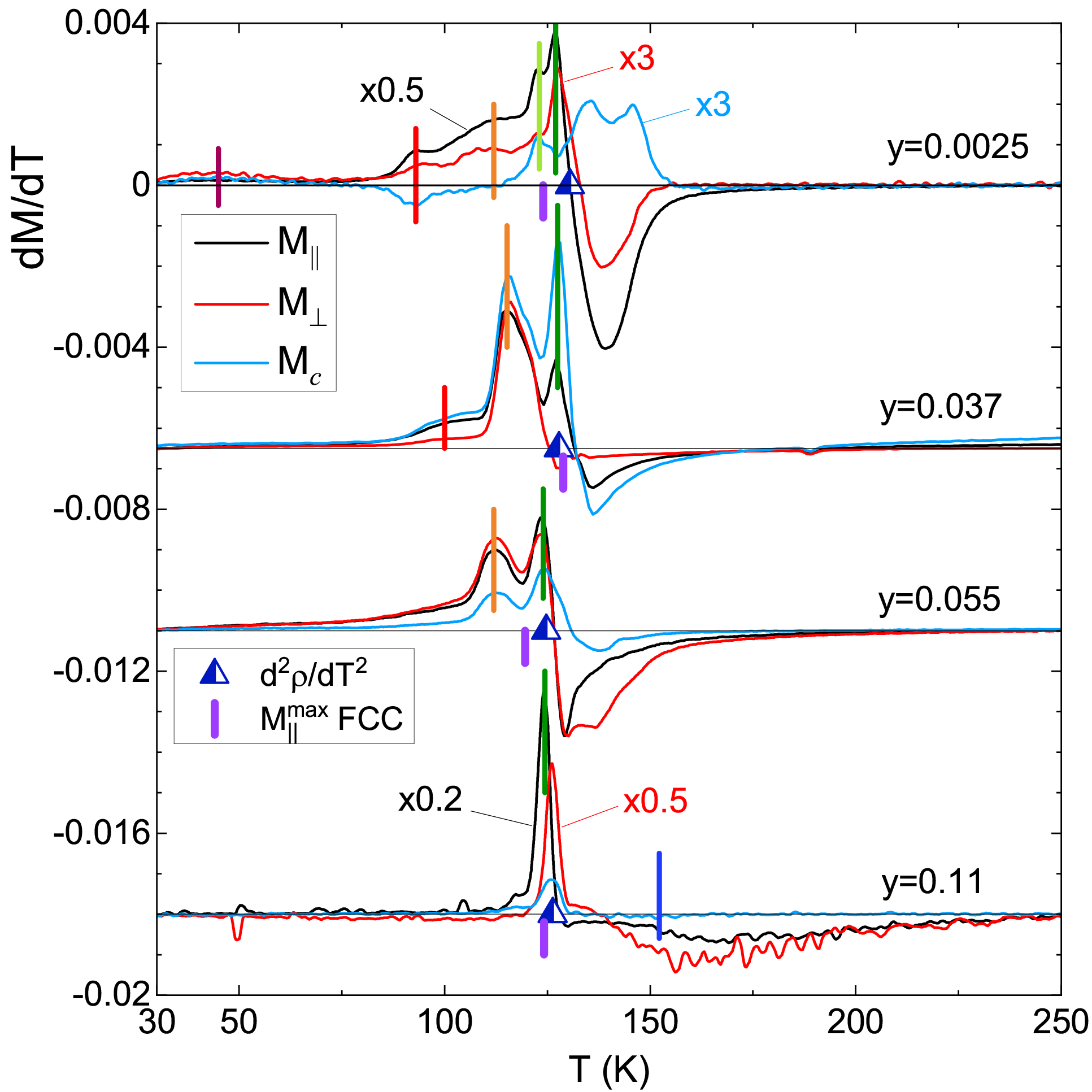}
\caption{(Color online) d$M$/d$T$ versus $T$ for different components of magnetization, $M_{\parallel}$ (black), $M_{\perp}$ (red), and $M_c$ (blue), measured in ZFC mode, for samples with different $y$. For better visibility the data for some components are multiplied by a factor shown in the figure; in addition, the plots for $y$ equal to 0.037, 0.055, and 0.11 are shifted downwards. Vertical lines (brown-to-green) mark several extrema of the derivatives, and blue line marks the position of shoulder on lowest graph. Resistive anomalies and maxima of FCC $M_{\parallel}$ are marked by blue triangles and purple bars, respectively.}\label{DerivM}
\end{figure}

The difference in the shape of the peaks is visualized by the plot of the derivative d$M$/d$T$ versus $T$, which we show in Fig.\ref{DerivM} for all ZFC components for several crystals, with resistive anomalies and maxima of FCC curves marked by blue triangles and purple bars, respectively, at the bottom of each plot. We observe that in all crystals with $y < 0.06$ there are precursors to the maximum of the peak, situated at lower temperatures, which we mark by vertical color lines; the maximum of the peek at highest $T$ is marked by green line. The d$M$/d$T$ shows most complicated structure in the crystal with smallest $y=0.0025$, starting from low $T$ just above 30 K, and extending towards higher temperatures, with several precursors, well correlated for different components of $M$ (some maxima for in-plane components are accompanied by minima for out-of-plane component, what signals spin reorientation). As $y$ increases the number of precursors decreases, until in case of crystal with $y=0.11$ there are no precursors left; instead, only one maximum is seen (marked by green line), accompanied by a shoulder at $T=152$ K (blue line).

Furthermore, we stress that the purple bar, which marks the maximum of the FCC curve, is always situated close to the maximum of the ZFC peak, indicated by green line. This strongly suggests that the maximum of ZFC peak may be associated with the FM component of the transition; the temperature of this component remains almost constant with the change of $y$. Conversely, the low-$T$ precursors in all cases with $y < 0.06$, and the shoulder in case of $y=0.11$ crystal, are most likely related to AFM components; as $y$ increases, the low-$T$ precursors are progressively wiped out, and only shoulder remains.

The complex structure of d$M$/d$T$ in case of least-doped crystal suggests series of consecutive phase transitions, accompanied by spin reorientations. Such behavior is consistent with the presence of disordered magnetic domains of various lateral sizes, which appear as a result of short-range magnetic ordering. Short-range ordering explains different influence of strong magnetic field and Ni-doping on the peak feature. While strong magnetic field destroys short-range ordering by orienting magnetic moments along the field, doping with Ni does not affect local magnetic ordering away from impurities, except for enhancing magnetic moments due to carrier scattering.

Trying to guess the types of local magnetic orderings, we recall that the system of coexisting localized and itinerant spins is governed by close competition between two interactions, the AFM superexchange coupling between localized spins, and the double-exchange FM coupling due to interaction between itinerant carriers and localized spins; theory shows that depending on material parameters such competition may produce different types of orderings, AFM-type (collinear, bicollinear, or checkerboard), all very close in energy, or FM-type \cite{Yin2010}. While long-range AFM collinear ordering is observed in undoped pnictides \cite{Cruz2008,Huang2008}, in Fe$_{1+\delta}$Te AFM ordering is bicollinear \cite{Bao2009}, accompanied by displacement of some Fe and Te ions leading to formation of FM-coupled Fe-Fe zig-zag chains with metallic conduction \cite{Fobes2014}. The modeling of inelastic neutron scattering in FeTe$_{1-x}$Se$_x$ reveals even more complicated picture of dynamical magnetic correlations; it requires replacement of single localized spin by FM- or AFM-coupled clusters of 4 NN Fe-spins, which interact with other clusters by two types of AFM correlations, collinear (stripe), or bicollinear \cite{Tranquada2020}. While bicollinear-coupled FM clusters are found at $x=0$ \cite{Zaliznyak2011}, at $x \simeq 0.5$ stripe-coupled AFM clusters dominate at low $T$, and bicollinear-coupled AFM clusters dominate at high $T$ \cite{Xu2016}; doping with Ni gradually eliminates stripe-coupled clusters, and drives the system towards bicollinear-coupled AFM clusters \cite{Xu2014}.

Given such complicated picture it is not at all surprising that we find peaks with two components, AFM- and FM-type. Accordingly, we suppose that in our mixed crystals 2 different types of AFM magnetic domains are formed, with prevailing stripe-coupled AFM clusters at low $T$, and bicollinear-coupled AFM clusters at higher $T$, which collectively give rise to series of consecutive phase transitions observed as precursors in the least-doped crystal ($y=0.0025$). Since with increasing Ni doping stripe correlations are suppressed, the low-$T$ precursors are wiped out, leaving magnetic domains of bicollinear type, which give rise to a shoulder at $y=0.11$.

It is more difficult to propose possible scenario for FM component to phase transition, which remains fixed at $T \approx 125$ K, almost unaffected by Ni doping. While we expect that some FM clusters may exist in our crystals, similar to situation observed in $x=0$ experiments \cite{Zaliznyak2011}, the coupling between clusters in the latter case is due to weak AFM-bicollinear interaction, which cannot be responsible for formation of FM magnetic domains. Instead, we suggest that delocalization of $d_{xy}$ orbital in our crystals with $x=0.35$ promotes increase of the density of itinerant carriers below $T \approx 125$ K, what enhances double-exchange interaction and likely favors local FM order. It is possible that local strains, induced by different size of Te and Se ions, contributes additionally to this development. Since $d_{xy}$ orbital survives Ni doping, the FM component is mostly unaffected by the increase of $y$, in stark contrast to AFM component, which changes. For further confirmation of this scenario investigation with local probes are necessary, what is beyond present study.

Based on the above analysis we conclude that the main origin of the anomalies observed in magnetization is the incoherent-coherent transition, which increases the density of itinerant carriers at low $T$, leading to change of $M$ with temperature, and to short-range magnetic order.

\subsection{Angle-dependent magnetoresistance}

Now we turn attention to anisotropic magnetoresistance (AMR), which may provide information on the anisotropy od spin correlations. The AMR is a phenomenon in which resistivity depends on the relative angle between directions of $I$ and $M$ \cite{Gorkom2001,Kokado2015,Ritzinger2023}. It results from the combination of spin-dependent scattering of carriers, and spin-orbit interaction, which induces mixing of spin-up and spin-down $d$ states; the mixing, which depends on $M$ direction, determines the density of unoccupied $d$ states at the Fermi level, and therefore affects carrier scattering.  Experiments show two basic classes of the AMR, noncrystalline AMR, with 2-fold symmetry, observed in polycrystalline materials, and crystalline AMR, which may include higher symmetry terms, reflecting magnetocrystalline anisotropy (MCA) of the material, that is, the dependence of $M$ on crystallographic directions. Microscopic theoretical models are rather involved, requiring knowledge of the band structure and of the scattering matrix. Experiments are usually interpreted based on phenomenological models, which consider different symmetry terms contributing to the AMR; we follow this method here.

The AMR in crystalline magnetic metals has been observed in many FM, AFM, or ferrimagnetic crystals and films \cite{Bason2009,Li2010,Wong2014,Gong2018,Miao2021}. At lowest temperatures it usually reflects the MCA, provided the external magnetic field is high enough so that spins are nearly aligned with the field. For crystals with tetragonal symmetry the presence of 4-fold symmetry of the MCA (biaxial anisotropy) is expected due to 4-fold symmetry of the lattice, while 2-fold symmetry terms (uniaxial anisotropy) appear on decreasing $T$ due to built-in strains and other factors \cite{Kokado2015,Li2010}. In case of the IBS materials the in-plane AMR has been described in pnictides \cite{Chen2008,Wang2009,Chu2010}, in FeSe \cite{Yuan2016}, and in several chalcogenide systems, including FeTe$_{1-x}$Se$_{x}$, with $x$ = 0, 0.39, and 0.94 \cite{Liu2021}. These studies, which were limited to non-SC compositions, observed 2-fold symmetry of the AMR below some characteristic temperatures, which in some cases could be identified, for example, as AFM ordering temperature in parent FeTe, nematic transition temperature in FeSe, or spin density wave transition or/and AFM ordering in pnictides. The origin of the 2-fold symmetry is linked to the prevailing anisotropic magnetic correlations existing at low temperatures \cite{AMRsymmetry}.

\begin{figure}
\centering
\includegraphics[width=8.5cm]{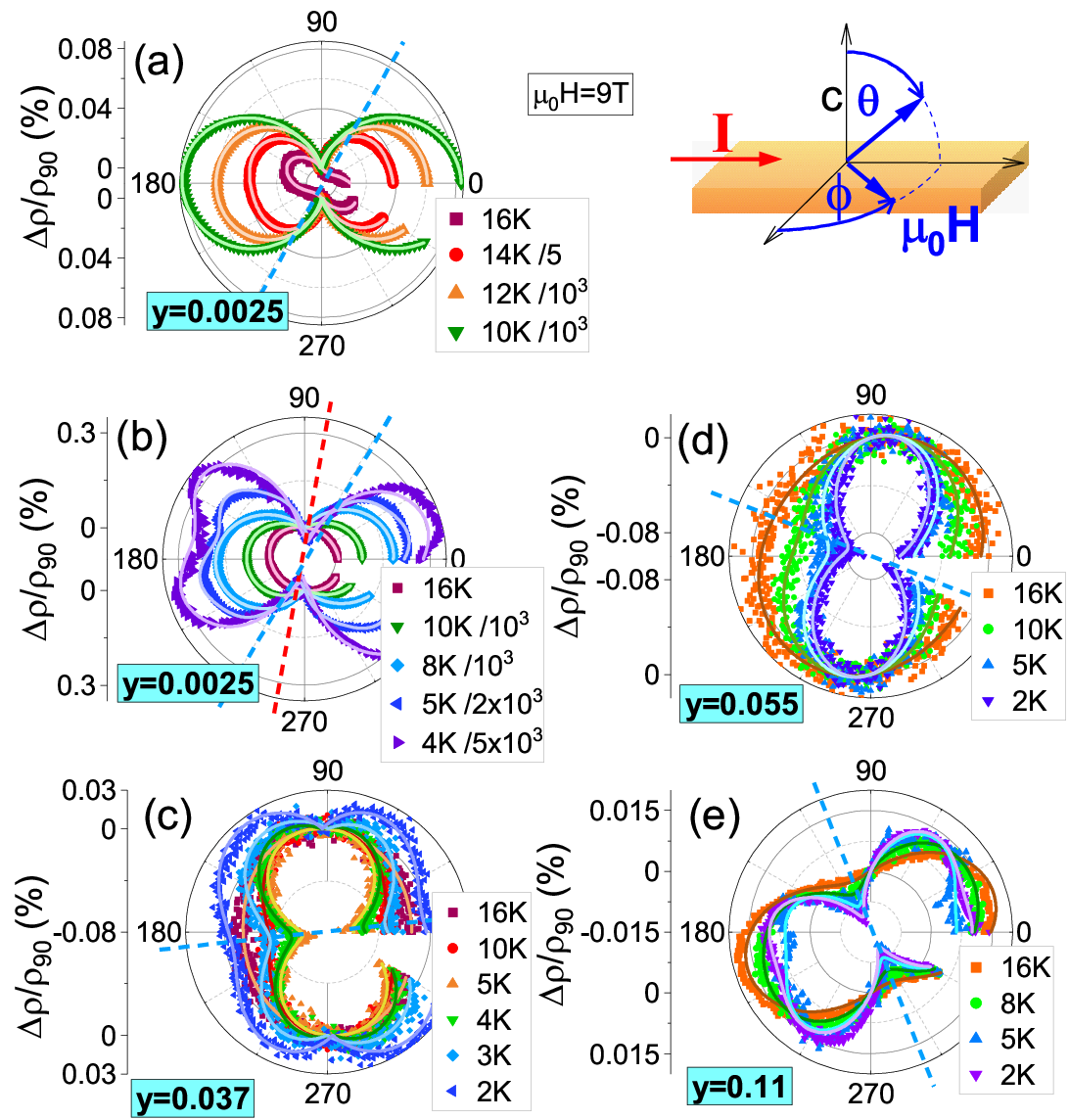}
\caption{(Color online) $\Delta \rho /\rho_{90}$ versus angle on planar plots ($\phi$ scans) for crystals with various $y$, 0.0025 (a-b), 0.037 (c), 0.055 (d), 0.11 (e). The points are experimental data, measured at magnetic field of 9 T for different temperatures labeled in the figures, and lines show the fitted dependencies as described in the text. In (a-b) the low-$T$ data are divided by a factor specified in the label. The inset at the top right shows configuration of the $\theta$ and $\phi$ scans.}\label{AMR}
\end{figure}

As we are interested in properties of Ni-doped crystals in a broad range of doping, it is important to evaluate the behavior of the AMR starting from SC compositions. In all samples studied in this work the magnetoresistance (MR) measured in the perpendicular magnetic field in the normal state starts to acquire substantial, negative values at temperatures below about 30 K. The positive conventional MR is negligible in this $T$-range. In SC samples below the $T_c$ the positive component of the MR starts to appear, related to suppression of superconductivity. In order to separate clearly the SC and non-SC components of the MR, we have measured the AMR for two scans illustrated in the inset to Fig.\ref{AMR}: $\phi$ is the azimuthal angle for rotation around $c$-axis, with $\phi$=0 for $H \perp c \perp I$, and $\theta$ is the angle between $c$ axis and $I$ direction, with $\theta$=0 for $H \parallel c \perp I$ (we have verified that Lorentz-force effects are insignificant in $\theta$ scans). Since the effect of the magnetic field on the SC component of the AMR should be minimized for $H \parallel I$, i.e. $\theta$ = $\phi$ = 90 deg, we normalize the AMR to resistivity $\rho_{90}$ measured at 90 deg, $\Delta \rho /\rho_{90} = (\rho - \rho_{90} )/\rho_{90}$, where $\rho$ is the resistivity measured at any given $T$, $H$, and $\theta$ or $\phi$.

$\Delta \rho /\rho_{90}$ for $\phi$ scans measured at various temperatures and magnetic field ${\mu}_0 H = 9$ T is shown in Fig.\ref{AMR} for several crystals with different $y$ (for discussion of $\theta$ scans see Appendix B). In the normal state ($T = 16$ K) the AMR shows dominant 2-fold symmetry in each crystal, with the positions of AMR minima depending on $y$, as indicated by dashed blue lines. In the crystal with $y=0.0025$ [Fig.\ref{AMR}(a-b)] the minimum is located at $\phi \simeq 60$ deg, that is, it is shifted away from either of the in-plane axes, towards nearest-neighbor (NN) Fe-Fe direction (but not exactly along it); in case of intermediate $y$ (0.037 and 0.055) it shifts closer to the vicinity of $\phi = 0$, i.e. to next NN (nNN) Fe-Fe direction; finally, at highest $y = 0.11$ the position is again closer to NN direction. Note that in case of intermediate $y$ the AMR maximum is located close to $\phi = 90$, confirming large scattering of carriers for this magnetic field orientation, as observed in magnetization data (Fig.\ref{Susc}).

Focusing first on the crystal with $y = 0.0025$, on decreasing $T$ below the $T_c$, down to 10 K, we observe rapid growth of positive $\Delta \rho /\rho_{90}$, with sharp minimum at $\phi$ = 90 deg and maximum at $\phi$ = 0, as expected in the SC state [Fig.\ref{AMR}(a)]. However, on further decrease of temperature, when the resistance becomes very small deeper in the SC state, the symmetry of the AMR changes into 4-fold [Fig.\ref{AMR}(b)], and the minimum of the AMR shifts away from $\phi = 90$ to $\phi = 80$ deg, as marked by red dashed line. Apparently, at this low $T$ the AMR due to suppression of the SC state becomes small, and magnetic component, related to the MCA, dominates.

When the doping with Ni increases to $y = 0.037$, the SC-related term is quenched, so that at 9 T it becomes visible only at $T < 4$ K as a small dip at $\phi = 90$ deg [Fig.\ref{AMR}(c)]. As the Ni doping increases further the SC term is no longer observable, so that pure MCA term is seen down to 2K [Figs.\ref{AMR}(d-e)]. Notably, in the two cases with intermediate $y$ (0.037 and 0.055) the dominant 2-fold symmetry of the MCA persists with unchanged position of the minimum for all temperatures. However, in case of largest $y$ [Figs.\ref{AMR}(e)] the 4-fold MCA becomes apparent at lowest $T$.

In order to quantify these observations, we fit the data by sum of terms which describe different angle dependencies expected for the SC and the MCA components: $S(x) = S_0 |\cos{(x)}|$, and $M_n (x) \sim M_n  \left[ \cos{[n (x - \frac{\pi}{n} - x_n)]} \right]$, respectively. Here $M_n$ and $x_n$ are amplitudes and phases of the MCA terms of $n$-fold symmetry, treated as fitting parameters ($x_n$ are almost constant for different $T$), while $S_0$ is the amplitude of SC term, extracted using boundary conditions from fitted $M_n$'s. We find that two MCA terms, $n=2$ (uniaxial anisotropy) and $n=4$ (biaxial anisotropy), are enough to fit the data. The fitted dependencies describe the data very well, as shown by continuous lines on Fig.\ref{AMR} (for more details see Appendix B). Both $M_2$ and $M_4$ amplitudes increase with decreasing $T$, reflecting growing influence of the MCA on scattering of carriers. However, while the $M_2$ amplitude is found to be finite for all dopings and temperatures, indicating that 2-fold symmetry term is dominant in most of the crystals, the $M_4$ amplitude varies. It reaches the largest magnitude at low $T$ in SC crystals, while it is suppressed at intermediate $y$, suggesting the development of spin nematicity with doping.

The origin of dominant 2-fold symmetry in FeTe$_{1-x}$Se$_x$ crystals may be traced to the nature of dynamic magnetic correlations of stripe and/or bicollinear types \cite{Tranquada2020,Zaliznyak2015,Xu2016}. Since these correlations break $C_4$ symmetry, therefore, taken separately, they should lead to uniaxial MCA. Our result shows, however, a mixed 2-fold and 4-fold symmetry, with the largest 4-fold component present in SC sample at lowest $T$ [Fig.\ref{AMR}(b)]. It is very likely that intrinsic disorder in our crystals (due to Te/Se and Fe/Ni mixture) leads to the coexistence of domains of limited lateral size with prevailing stripe or bicollinear correlations. The AMR experiment averages over them, producing mixed symmetry, and the increasing contribution of stripe domains on decreasing temperature leads to 4-fold component most pronounced at lowest $T$. Note that cases of stripe-type spin-density wave orders, which preserve $C_4$ symmetry, called spin-charge density wave or spin-vortex crystal \cite{Fernandes2022}, have been observed in some pnictides in limited regions of phase diagrams \cite{Allred2016,Meier2018}, however, there are no reports of such observations in FeTe$_{1-x}$Se$_x$.

In order to visualize the evolution of the MCA symmetry with doping and temperature we create a map of the ratio $M_4 /M_2$, and we use it as a background in a phase diagram, discussed in the next section.

\section{Phase diagram}

\begin{figure}
\centering
\includegraphics[width=7.5cm]{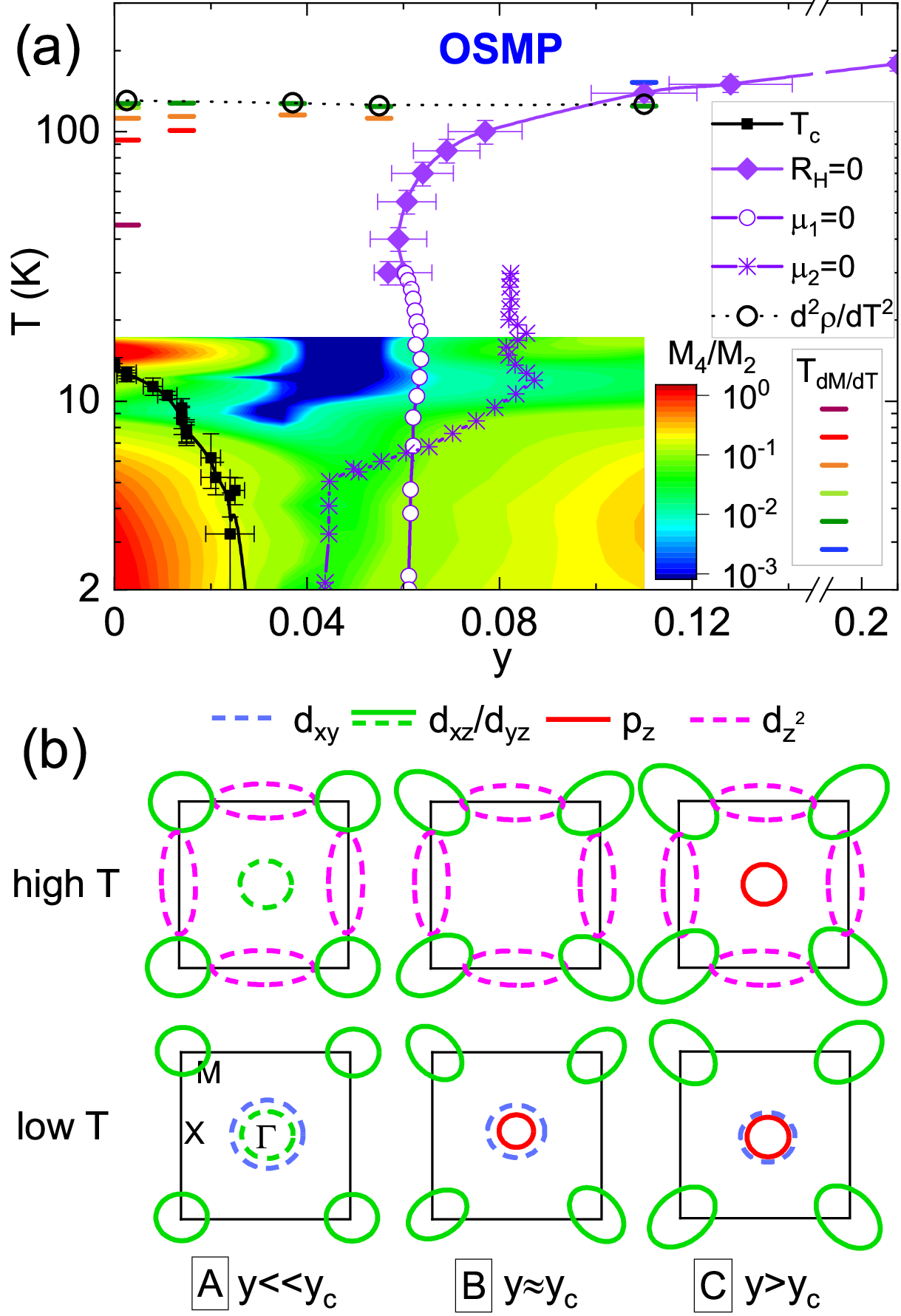}
\caption{(Color online) Phase diagram of Fe$_{1-y}$Ni$_y$Te$_{0.65}$Se$_{0.35}$. Color background shows the ratio $M_4 /M_2$ extracted from AMR. Points indicate temperatures: $T_c$ (black squares), $R_H =0$ (diamonds), ${\mu}_1 = 0$ (open circles), ${\mu}_2 = 0$ (stars), resistive anomalies (black spheres), $T_{dM/dT}$ - peak positions of $dM/dT$ (colors as in Fig.\ref{DerivM}). All lines are guides to the eye. (b) Schematics of the band structure at the Fermi level at high $T$ (top raw) and at low $T$ (bottom raw) for (A) $y \ll y_c$, (B) $y \approx y_c$, and (C) $y > y_c$; continuous and dashed lines indicate electron-like and hole-like bands, respectively.}\label{Diag}
\end{figure}

We summarize all experimental results on the phase diagram in Fig.\ref{Diag}(a) (with $T$ on a logarithmic scale). The map of the ratio $M_4 /M_2$, obtained from the AMR, is plotted as a background in a lower portion of the figure. The red indicates presence of substantial $M_4$ terms, reaching about 1 at $ y\approx 0$ at lowest $T$. Green represents area with $M_4 /M_2$ suppressed down to between 0.01 and 0.1, while dark blue area specifies regions of strictly twofold symmetry of the MCA. The map reveals peculiar dependence of $M_4 /M_2$ on temperature and doping: there is a clear suppression of $M_4 /M_2$ for intermediate $y$, but, in addition, there is suppression in the vicinity of $T = 10$ K for all doping levels. We add to the graph all data obtained from transport experiments, displayed previously in Fig.\ref{TDiag}: the $T_c$, the $R_H$-line, ${\mu}_i$-lines, and the temperatures of the resistive anomalies. Finally, we also mark by color horizontal bars the temperatures $T_{dM/dT}$, at which $dM/dT$ shows extrema, using the same colors as lines in Fig.\ref{DerivM}.

The phase diagram shows two important features. One of them is seen in high-$T$ region, and involves the correlation between resistive anomaly and the magnetization anomaly, both approaching $R_H$-line on increasing $y$. This correlation suggests that both anomalies occur as a result of the transition into OSMP phase, in which $d_{xy}$ orbital becomes incoherent. Interestingly, the temperature at which the transition occurs depends only weakly on $y$, what underscores the fact that the $d_{xy}$ orbital survives Ni doping. In Fig.\ref{Diag}(b) in the bottom raw we re-plot schematically the band structure at the Fermi level at low $T$ based on ARPES experiments \cite{Rosmus2019,Yi2015,Huang2022}, as discussed previously, for (A) $y \ll y_c$, (B) $y \approx y_c$ and (C) $y > y_c$. In the top raw the band structure in the OSMP phase is shown: in all diagrams hole pocket of $d_{xy}$ origin is washed away but instead hole pockets at X point, of $d_{z^2}$ origin, appear. Note that in (C) the electron-like pocket of $p_z$ origin is likely surviving at high $T$, contributing to the upward shift of the $R_H$-line with increasing $y$.

Previously, the appearance of negative component to $R_H$ at low-$T$ has been interpreted as evidence of incoherent-coherent transition in magnetotransport of Te-annealed Fe$_{1+y}$Te$_{1-x}$Se$_x$ crystals \cite{Otsuka2019,Jiang2023}. Interpretation of the linear versus magnetic field dependence of Hall resistivity, observed in that case, requires considerable assumptions. As suggested in Ref. \cite{Huang2022}, a possible origin of negative $R_H$ is the influence on the Hall coefficient of vertex corrections due to spin fluctuations, provided the nesting conditions for spin fluctuations are fulfilled \cite{Fanfarillo2012}. In this respect, our experimental findings avoid any complications in interpretation, because we are able to show convincingly the electron doping at large $y$, based on non-linear dependence of Hall resistivity on magnetic field. Moreover, the nesting conditions are not fulfilled at large $y$, therefore vertex corrections should be suppressed \cite{Fanfarillo2012}. Thus, the reemergence of positive $R_H$ at high temperatures in crystals with large $y$, in which the only hole band at low $T$ is of $d_{xy}$ origin, seems to be strong confirmation of the contribution to transport of $d_{z^2}$ holes originating from pockets at X points in the OSMP phase.

The second interesting feature is the dependence of the $M_4 /M_2$ on $y$ and $T$, displayed in the low-$T$ portion of the phase diagram. Considering first $y$-dependence at the lowest $T$ we recall that the gradual replacement of stripe correlations by bicollinear correlations is expected with increasing $y$ \cite{Xu2017}, what would eliminate the domains with stripe correlations, leading to the reduced magnitude of $M_4$. This, together with doping-induced strains most likely leads to domination of 2-fold AMR component at intermediate $y$. On further increase of $y$ up to 0.11, weak 4-fold component reappears. We believe that this is of different origin than the 4-fold term at small $y$, since at this large Ni content stripe correlations are already absent. Rather, it is a consequence of the localization of electron carriers and enhanced anisotropy of magnetization at low $T$, what probably induces biaxial MCA.

More puzzling is the non-monotonic dependence of the $M_4 /M_2$ on temperature, with obvious suppression around 10 K. Note that at low $y$ this suppression seems to correlate with slightly stronger suppression of the $T_c$ at about 8-10 K. This correlation is further confirmed by the suppression of the amplitude of SC component of the AMR in exactly the same $T$-range (see Fig. \ref{A3} in Appendix B). This observations may corroborate the connection between local nematicity and the suppression of superconductivity, observed by recent quasiparticle scattering experiments \cite{Zhao2021}.

Even more interesting is the comparison between the map of $M_4 /M_2$ and the ${\mu}_2$-line. It appears that the ${\mu}_2$-line encircles the region of suppressed $M_4$ on the side of large $y$. We recall that the unusual dependence of ${\mu}_2$-line on $y$ and $T$ originates from the switching from electron-like mobility at low $T$ to hole-like mobility at high $T$ in the vicinity of $y_c$, as illustrated in Fig.\ref{LowHall}. This means that the suppressed $M_4$ region is likely associated with the presence of hole-like band at the Fermi level, which at $y \gtrsim y_c$ is of $d_{xy}$ origin. It is tempting, therefore, to link spin-nematicity described here to the behavior of $d_{xy}$ orbital. It is worth to mention that many recent studies of FeSe find that nematicity in this system is not restricted to $d_{xz}$ and $d_{yz}$ orbitals, but involves $d_{xy}$ orbital as well, in fact, $d_{xy}$ may be crucially important for the development of the nematic state \cite{Rhodes2022,Li2020,Rhodes2021}. Our results suggest that the importance of $d_{xy}$ orbital is not restricted to FeSe, but it may be a more general phenomenon. What exactly is the role of $d_{xy}$ orbital for spin-nematicity cannot be judged on the basis of present experiment, but further detail studies of Ni-doped FeTe$_{1-x}$Se$_x$ may turn out to be helpful for understanding of this role.

\section{Summary}

In summary, using systematic magnetotransport (resistivity, Hall effect, AMR) and magnetization measurements we have determined the phase diagram of single crystalline Fe$_{1-y}$Ni$_y$Te$_{0.65}$Se$_{0.35}$ ($0 < y < 0.21$), in which Ni substitution dopes the system with electrons, eliminating some of the hole pockets from Fermi level, and leaving only one, originating from $d_{xy}$ orbital.

We show that doping suppresses superconductivity to zero at $y \approx 0.03$, and induces transition from hole-dominated into electron-dominated conduction, with a boundary defined by Hall coefficient equal zero. At low temperature the boundary runs almost parallel to $T$-axis at value of $y \approx 0.06$, but at high temperatures it terminates the electron-dominated region in the vicinity of $T \approx 125 \div 178$ K, indicating reversal of the conduction back into hole-dominated at higher $T$. Anomalies in magnetization and resistivity are observed at temperatures which approach high-$T$ boundary of the electron-dominated region. Analysis of these effects suggests a link with the incoherence-coherence transition due to localization of $d_{xy}$ orbital on increasing $T$, and, related to it, appearance of the $d_{z^2}$ hole pockets at X points of the Brillouin zone in the OSMP phase, as recently uncovered by ARPES \cite{Huang2022}. Thus, this result appears to be the first unambiguous observation of the OSMP from magnetotransport measurements. The low-$T$ AMR shows mixed 4-fold and 2-fold rotational symmetry of in-plane magnetocrystalline anisotropy, with the 4-fold term the largest at small $y$, and suppressed at intermediate $y$. These results are consistent with the mixed stripe/bicollinear magnetic correlations at small $y$, and suppression of stripe correlations at intermediate $y$, indicating development of spin nematicity with increasing Ni doping, likely contributing to the suppression of superconductivity.

\section*{Acknowledgments}

The work has been supported by Polish NSC Grant No. 2014/15/B/ST3/03889. The research was partially performed in the laboratory co-financed by the ERDF Project NanoFun POIG.02.02.00-00-025/09.

\appendix

\section{Evaluation of carrier mobilities}

\begin{figure}
\centering
\includegraphics[width=5.5cm]{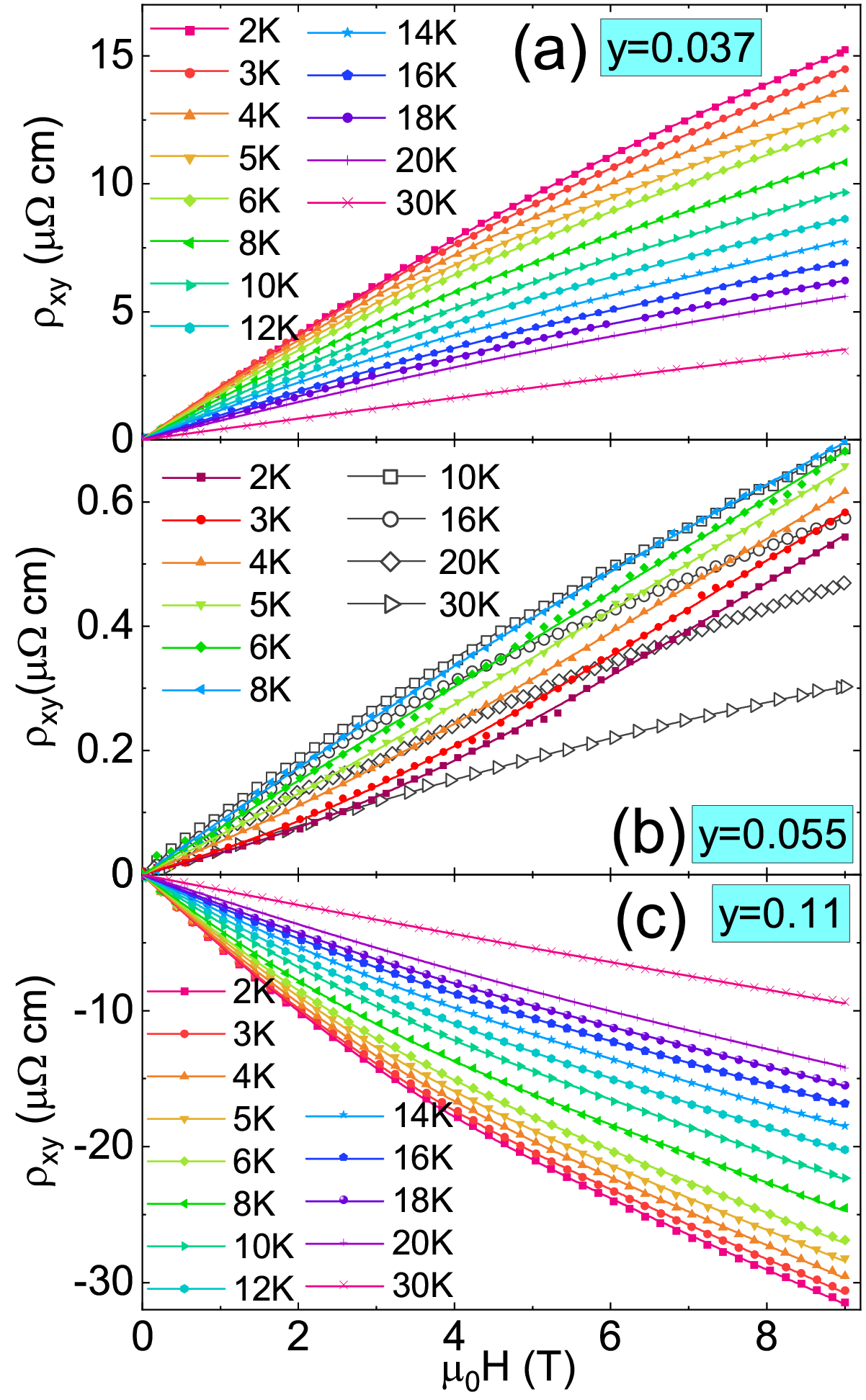}
\caption{(Color online) $\rho_{xy}$ versus $\mu_0 H$ for $y = 0.037$ (a), 0.055 (b), and 0.11 (c). The points are experimental data, and lines are fitted dependencies (A1).}\label{A1}
\end{figure}

Fig.\ref{A1} shows the dependence of Hall resistivity, $\rho_{xy}$ on the magnetic field for some of the crystals. In the case of each sample the dependence evolves from near linear at 20-30 K to nonlinear at 2 K. The nonlinearity is, however, dependent on $y$. Specifically, the $\rho_{xy} (H)$ evolves from positive values and convex shape in Fig.\ref{A1}(a) to negative values and concave shape in Fig.\ref{A1}(c), suggesting that two types of carriers are involved in each case, with the same polarity but different mobilities, positive in the former case, and negative in the latter. In case of sample with intermediate doping, shown in Fig. \ref{A1}(b), the $\rho_{xy}$ is still positive, but there is a change of shape from convex for $T > 6$ K to concave for $T < 6$ K, signaling a change of sign of minority carriers from positive to negative.

In order to extract carrier concentrations and their mobilities, we fit the Hall resistivity data using the exact expressions derived from the matrix formalism, which has been proposed in the past for the description of multicarrier semiconductor systems \cite{Kim1993,Kim1999}. In this formalism one type of carriers is understood as collection of all carriers with the same mobility. While in principle the formalism allows to consider infinite types of carriers, in practice the fit must be limited to three types, and, in fact, fitting the 3-carrier expressions is usually difficult. For 2-carrier system with the carrier concentrations $n_i$ and mobilities $\mu_i$ ($i=1,2$) the Hall resistivity is given by (more complicated 3-carrier expressions may be found in Ref.\cite{Kim1999}),
\begin{flalign}
\rho_{xy} &= {\rho} ({\mu_0} H) [{\alpha}+ {\beta_1}({\mu_0 H})^2]/[1+{\beta_2}({\mu_0 H})^2],&\\ \nonumber
\alpha &= f_1 \mu_1 + f_2 \mu_2, \,\quad \beta_1 = (f_1 \mu_2 + f_2 \mu_1) {\mu_1}{\mu_2},& \\ \nonumber
\beta_2 &= {(f_1 \mu_2 + f_2 \mu_1)}^2, \,\quad f_1 + f_2 = 1.  \nonumber
\end{flalign}
In these equations mobilities are positive (negative) for positive (negative) carriers, while the parameters $f_i$ are related to the concentrations of carriers by relation $n_i = f_i /(q \mu_i \rho)$, where $q$ is the carrier charge.

Data for most of the samples are very well fitted by the above expressions, using 3 fitting parameters ($f_1$, $\mu_1$, and $\mu_2$). The fitted curves are shown by continuous lines in Fig.\ref{A1}.

\section{Angular magnetoresistance}

In magnetic crystal the resistivity tensor $\rho_{ij}$ in the presence of magnetic field depends on the direction cosines,
$\alpha_i$, of the magnetization vector, and may be expressed as series expansions in ascending powers of $\alpha_i$ \cite{Bason2009,Li2010}

\begin{flalign}
\rho_{ij} (\alpha) &= \sum_{k,l,m,...=1}^{3} (a_{ij} + a_{kij}{\alpha}_k + a_{klij}{\alpha}_k {\alpha}_l &\\  \nonumber
&+ a_{klmij}{\alpha}_k {\alpha}_l {\alpha}_m + a_{klmnij}{\alpha}_k {\alpha}_l {\alpha}_m {\alpha}_n +...), \nonumber
\end{flalign}
where $i,j=1,2,3$ and $a$'s are expansion coefficients. Symmetry considerations for tetragonal crystal, and assumption of the current direction along one of the main crystal axis, reduces this equation to a simple expression for magnetoresistance
\begin{equation}
{\Delta}{\rho}/{\rho}_0 (x) = C_0 + \sum_{n} M_n \left[ \cos{[n (x - x_n)]}\right],
\end{equation}

where $x$ is the angle between magnetization vector and the main crystal axis, $C_0$ is a constant, and $M_n$ and $x_n$ are the amplitude and the phase of $n$-fold symmetry.

\begin{figure}
\centering
\includegraphics[width=8.5cm]{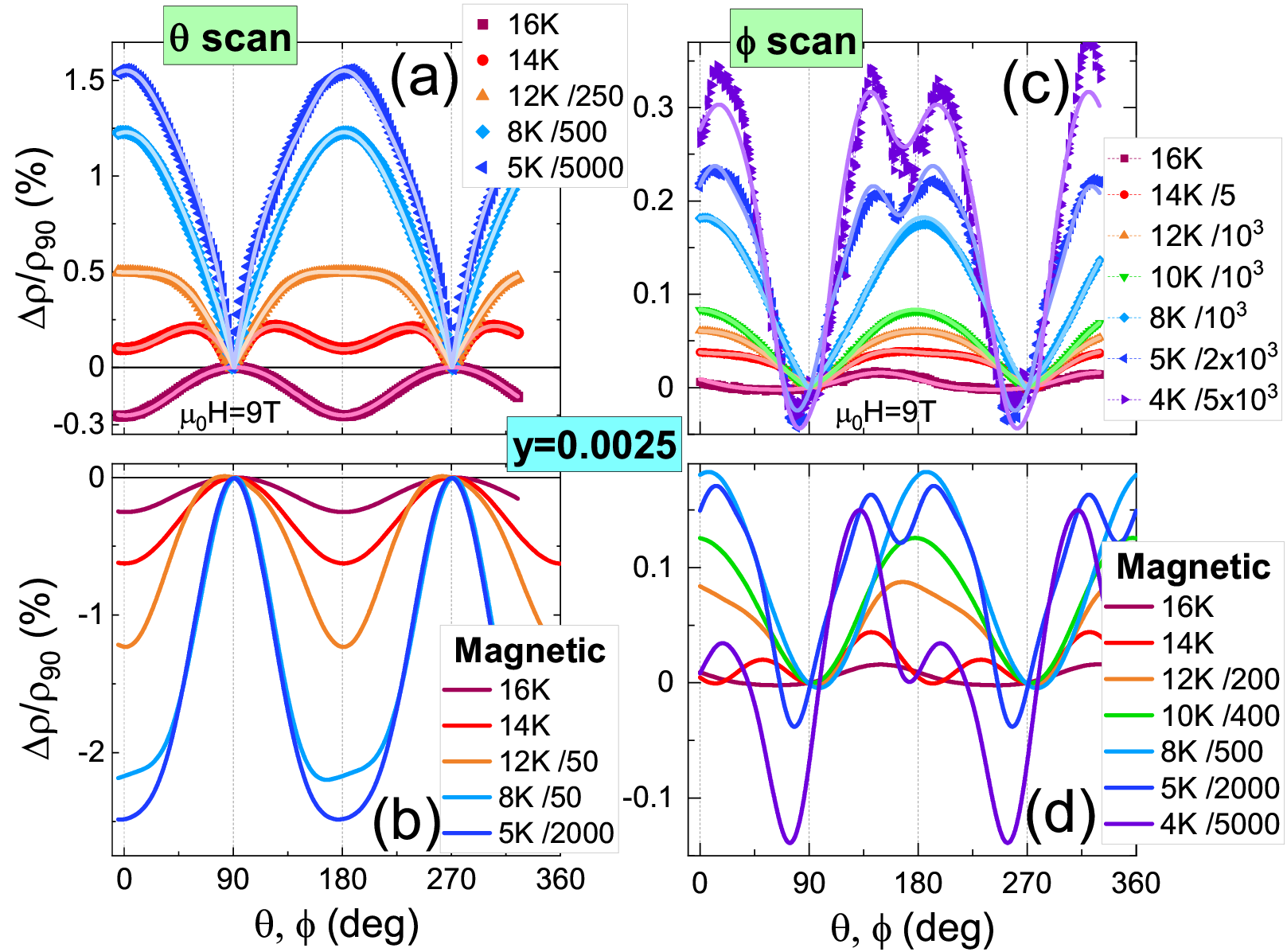}
\caption{(Color online) $\Delta \rho /\rho_{90}$ for crystal with $y = 0.0025$ versus angle, measured at magnetic field of 9 T and several different temperatures as labeled in the figure, for $\theta$ scan (a), and $\phi$ scan (c). The points are experimental data, and lines show the fitted dependencies (A1). The magnetic parts of the fitted dependencies are shown for $\theta$ scan in (b), and for $\phi$ scan in (d). The low-$T$ data in each figure are divided by a factor shown in the label.}\label{A2}
\end{figure}

Figs. \ref{A2}(a) and \ref{A2}(c) show the experimental data for sample with $y=0.0025$ measured for $\theta$ and $\phi$ scans, respectively. The data for the $\theta$ scan [Fig. \ref{A2}(a)] in the normal state, at $T = 16$ K, show that the AMR is negative, and has 2-fold symmetry, with minima located at $\theta$=0, i.e., in the out-of-plane direction. This is consistent with the picture of disordered cluster state with in-plane AFM ordering within clusters. Such clusters are source of substantial carrier scattering, which is reduced by out-of-plane magnetic field aligning the spins out-of-plane. Similar behavior is seen for all of the samples in the normal state. Decreasing $T$ below the $T_c$ results in the rapid growth of positive $\Delta \rho /\rho_{90}$, with the minimum at $\theta$ = 90 deg and maximum at $\theta =0$, as expected in the SC state. To display the low-$T$ data on the same graph, we divide it by large factors, as specified in the labels.

\begin{figure}
\centering
\includegraphics[width=8cm]{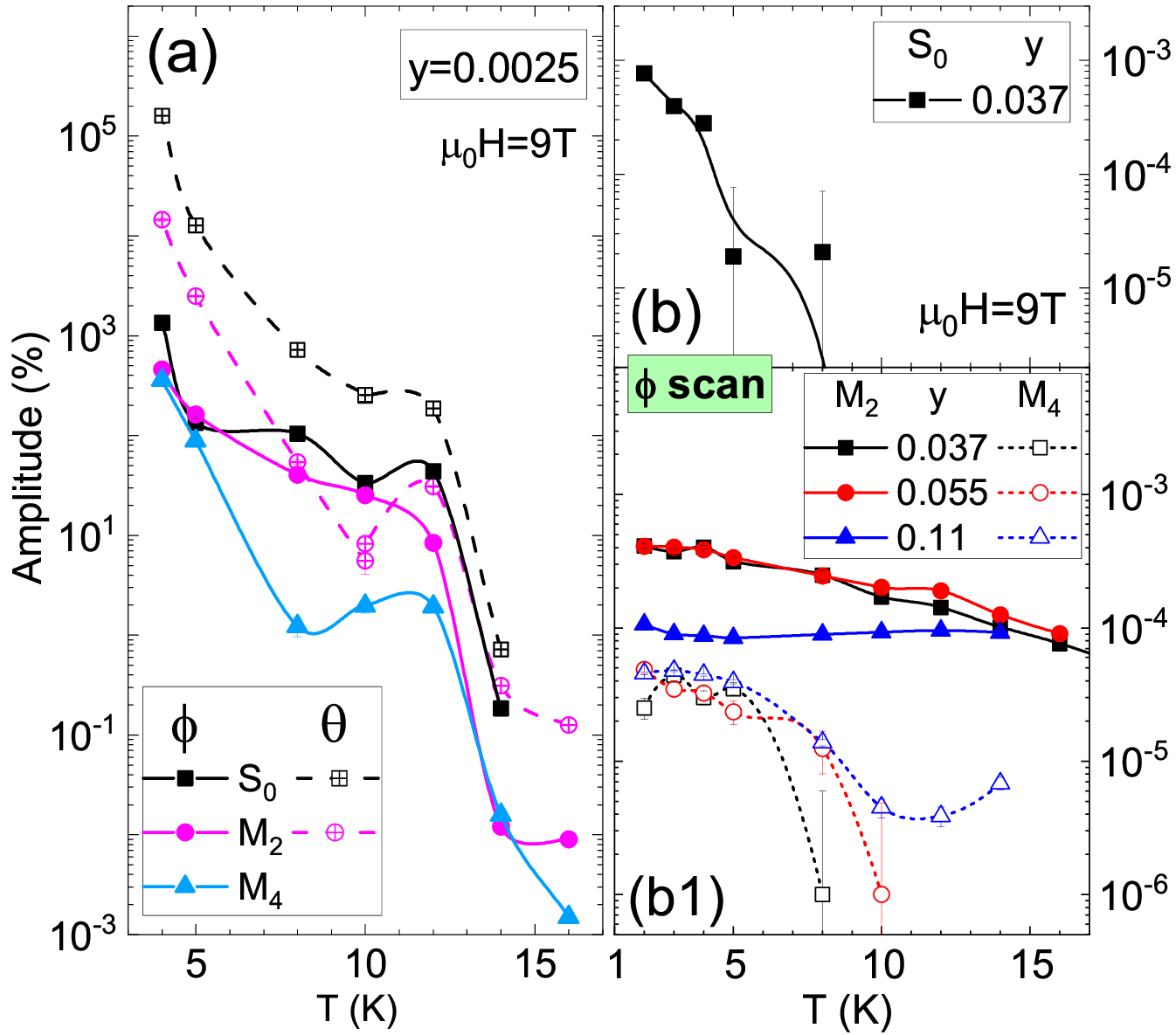}
\caption{(Color online) The amplitudes of superconducting ($S_0$) and magnetic ($M_2$ and $M_4$) components of the AMR versus $T$. (a) $y=0.0025$, $\phi$ scan (full points), and $\theta$ scan (open points). (b) $\phi$ scan, $S_0$ for $y=0.037$. (b1) $\phi$ scans for $y = 0.037, 0.055$, and 0.11, $M_2$ (full points) and $M_4$ (open points). All lines are guides to the eye.}\label{A3}
\end{figure}

$\Delta \rho /\rho_{90}$ for $\phi$ scan (Fig.\ref{A2}(c)) is about five times smaller than for $\theta$ scan. In the normal state ($T = 16$ K) it shows apparent 2-fold symmetry, with the minima of the AMR located at $\phi \simeq 60$ deg. As $T$ is decreased below the $T_c$ the behavior somewhat similar to the $\theta$ scan appears, i.e. $\Delta \rho /\rho_{90}$ grows the most in the vicinity of $\phi$ = 0, and the least for $\phi$ = 90 deg, indicating the suppression of SC state by magnetic field. However, at very low $T$, when the resistance becomes very small, the symmetry of the AMR changes into 4-fold, with the minimum of the AMR shifted away from $\phi = 90$ to $\phi = 80$ deg, clearly indicating that it is not related to SC state suppression. Instead, it reflects the minimum of magnetic component of the AMR, most likely related to the MCA.

In order to account properly for the superconducting ($S$) and magnetic ($M$) components of the AMR, we fit the experimental data by the following dependencies,
\begin{flalign}
{\Delta}{\rho}/\rho_{90} (x)&= S(x) + M(x),&\\ \nonumber
S(x) &= S_0 |\cos{(x-x_0 )}|, &\\  \nonumber
M(x) &= \sum_{n} M_n \left[ \cos{[n (x - \frac{\pi}{n} - x_n)]} + C_n \right],&\\ \nonumber
C_n & = \cos{[n (-{\pi}/2 - x_n + x_0)]}.  \nonumber
\end{flalign}
Here $M_n$ and $x_n$ are the amplitude and the phase of $n$-fold symmetry of the magnetocrystalline part, respectively; $x_n$ vary only slightly with the change of $T$, what simplifies the fits. $C_n$ are constants resulting from the normalization of AMR by $\rho_{90}$ (i.e. at $x = {\pi}/2$), and $x_0$ is a small phase shift of the SC component away from $\pi /2$ resulting from rotator misalignment. It has been shown that the AMR for tetragonal symmetry is well described by $n=2$ (uniaxial anisotropy) and/or $n=4$ (biaxial anisotropy) \cite{Li2010}. When doing the fits, we first determine ${\Delta}{\rho}/\rho_{90}$ at $x = 0$, and use it to eliminate $S_0$ from the equations; $M_n$, $x_n$ and $x_0$ are used as fitting parameters. This procedure results in excellent description of the data, as indicated by continuous lines in Figs.\ref{A2}(a) and \ref{A2}(c). The $M$ parts extracted from the fits, shown in Figs. \ref{A2}(b) and \ref{A2}(d), illustrate the growth of 4-fold symmetry at low $T$.

The $T$-dependencies of the amplitudes $S_0$ and $M_n$ are shown for several crystals in Fig.\ref{A3}. In the crystal with $y=0.0025$ both $S_0$ and $M_n$ amplitudes increase by many orders of magnitude with decreasing $T$ for both $\theta$ and $\phi$ scans (Fig.\ref{A3}(a)). In the limit of low temperatures the ratio between $M_4$ and $M_2$ for $\phi$ scan, $R = M_4 /M_2$, is about 1. However, it falls down to about 0.1 in the region between 8 K and 10 K, and again grows at higher $T$. Thus, the symmetry of the MCA changes with $T$. Interestingly, in the $T$-region with small $R$, which may be treated as indication of prevailing 2-fold symmetry, the $S_0$ amplitude seems somewhat suppressed. Turning to the crystals of intermediate Ni doping, $y$ = 0.037 and 0.055 (Fig.\ref{A3}(b1)), we observe that $R$ is about 0.1 at low $T$, and falls down to zero at higher temperatures, clearly confirming 2-fold symmetry.


\begin{thebibliography}{99}
\bibitem{AMRsymmetry}
It should be underscored that the 2-fold symmetry of the MCA, detected by the AMR, called "spin nematicity" induced by the magnetic field, is a signature of the anisotropy of spin correlations. It may not be necessarily identical with spontaneous electronic nematicity observed in the absence of the magnetic field, although in many cases it is observed below exactly the same temperature \cite{Liu2021}.
\end{thebibliography}
\end{document}